\documentclass[useAMS,usenatbib,usegraphicx]{mn2e}

\usepackage{color}
\newcommand\culine[1]{\bgroup \markoverwith {\textcolor{#1}{\rule[-0.5ex]{2pt}{0.4pt}}}\ULon}

\pagerange{\pageref{firstpage}--\pageref{lastpage}}
\pubyear{2011}

\newcommand{\be}{\begin{equation}}
\newcommand{\ee}{\end{equation}}
\newcommand{\ho}{\mathrm{Ho}}

\newcommand{\dZ}{\mathrm{dZ}}
\newcommand{\dNdz}[1]{\frac{dN^{#1}}{dz}}

\begin{document}

\title[An optimal estimator for the CMB-LSS angular power spectrum]
{An optimal estimator for the CMB-LSS angular power spectrum and its application to WMAP and NVSS data}

\author[F.~Schiavon et al.] 
{F.~Schiavon $^{1,2,3,4}$\thanks{E-mail: f.schiavon@ira.inaf.it}, 
F.~Finelli $^{3,4}$, A.~Gruppuso $^{3,4}$, A.~Marcos-Caballero $^{5,6}$, 
\newauthor P.~Vielva $^{5}$, R.~G.~Crittenden $^{{7}}$, 
R.~B.~Barreiro $^{5}$, E.~Mart\'inez-Gonz\'alez $^{5}$ \\
$^1$ Dipartimento di Astronomia, Universit\`a degli Studi di Bologna,
via Ranzani 1, I-40127 Bologna, Italy \\
$^2$ INAF-Istituto di Radioastronomia, via Gobetti 101, I-40129 Bologna, Italy\\ 
$^3$ INAF-IASF Bologna, Istituto di Astrofisica Spaziale e Fisica Cosmica 
di Bologna \\ via Gobetti 101, I-40129 Bologna, Italy \\
$^4$ INFN, Sezione di Bologna, Via Berti Pichat 6/2, I-40127 Bologna, Italy \\
$^5$ Instituto de F{\'\i}sica, (CSIC - Universidad de Cantabria), Avda. Los Castros s/n, E-39005, Santander, Spain \\
$^6$ Departamento de F{\'\i}sica Moderna, Facultad de Ciencias, Universidad de Cantabria, Avda. Los Castros s/n, \\ E-39005, Santander, Spain \\
$^7$ {Institute of Cosmology and Gravitation, University of Portsmouth, Dennis Sciama Building, Burnaby Rd.,}\\ {Portsmouth, Hampshire, PO1 3FX, UK}}

\label{firstpage}

\maketitle
\begin{abstract}
We use a Quadratic Maximum Likelihood (QML) method to estimate the angular power spectrum 
of the cross-correlation between cosmic microwave background and large scale structure maps as  
well as their individual auto-spectra. We describe our implementation of this method and demonstrate its accuracy on simulated maps. 
We apply this optimal estimator to WMAP 7-year and NRAO VLA Sky Survey (NVSS) data and explore the robustness 
of the angular power spectrum estimates obtained by the QML method. With the correction of the declination systematics in NVSS, we can  
safely use most of the information contained in this survey.
We then make use of the angular power spectrum estimates obtained by the QML method to derive
constraints on the dark energy critical density in a flat $\Lambda$CDM model by different likelihood prescriptions. 
When using just the cross-correlation between WMAP 7 year and NVSS maps with 1.8$^\circ$ resolution, 
the best-fit model has a cosmological constant of approximatively 
70$\%$ of the total energy density, disfavouring an Einstein-de Sitter Universe 
at more than 2 $\sigma$ CL (confidence level). 
\end{abstract}


\section{Introduction}

Understanding of the nature of dark energy is one of the outstanding questions in observational 
cosmology. Since the discovery of the present acceleration of the Universe by the measurement of the luminosity distance of distant 
type Ia supernov$\ae$ (SN Ia) \citep{riess,perlmutter}, several observations \citep[e.g.,][]{tegmark04,eisenstein,larson}
have converged to a cosmological concordance model in which 
an unknown component having a negative pressure density - `dark energy' - that contributes $\sim 2/3$ of the total energy budget of the Universe. 
At present the precise nature of dark energy, parameterised by its equation of state, can only be weakly constrained using a range 
of cosmological tests, but indications are that its behaviour is close to that expected from a cosmological constant.

A key strategy in determining the nature of dark energy is to combine as many 
different observations as possible, including the luminosity distances for Type Ia supernovae, the baryonic acoustic oscillation (BAO) 
scale observed in galaxy surveys, anisotropies of the cosmic microwave background (CMB) and weak lensing surveys. 
Cross-correlations among the above observations also contain precious cosmological information about dark energy.  
Ambitious space projects have been proposed to address the dark energy question with this strategy, 
including EUCLID \citep{euclid}, which will focus on 
BAO and weak lensing, and WFIRST \footnote{http://wfirst.gsfc.nasa.gov}, an infrared satellite with a focus yet to be specified.  
In the meantime, ground based programs such as DES \footnote{http://www.darkenergysurvey.org/}, PanSTARRS 
\footnote{http://pan-starrs.ifa.hawaii.edu/public/}, 
LSST \footnote{http://www.lsst.org/lsst/} will also improve the current understanding of structure formation and provide 
excellent galaxy surveys to cross-correlate with the CMB anisotropy maps from {\sc Planck} 
\citep{Planck}.

One of the key indicators of the presence of dark energy are CMB fluctuations created by the late Integrated Sachs-Wolfe (ISW) effect \citep{SW}. 
When the Universe is not completely matter dominated, CMB anisotropies are created at late times and these contribute most 
at large angular scales \citep{KS}.
Since the low multipoles of the CMB angular power spectrum are mostly affected by cosmic variance, an extraction of the ISW 
part solely from CMB data is rather difficult, but it is feasible when CMB is cross-correlated with large scale structure (LSS) 
\citep{CT1995}. Several positive detections of the ISW-LSS cross-correlation have been performed since the release of 
the WMAP first year data by using different tracers of LSS and statistical 
estimators 
(e.g. \cite{afshordi,boughn04,fosalba03,noltaetal,vielva,PBM,Ho,
Giannantonioetal2008,francispeacock}, \cite{dupe} and references therein).  

One of the purposes of this paper is to develop tools to estimate the angular power spectrum {(APS)} of the cross-correlation between 
CMB and LSS by a quadratic maximum likelihood {(QML)} method. The QML method in this context has a number of advantages: 
foremost, given the low signal-to-noise expected for the ISW-LSS cross-correlation, it is essential to use a minimum variance method, such as QML, to estimate the cross power spectrum.  In addition, based in pixel space,  the QML method is ideal for accounting for the incomplete sky coverage and masks of the surveys.  Finally, while the QML method is expensive computationally, the fact that the ISW signal is primarily at low multipoles means that it is tractable to constrain it on maps using only a modest resolution.    
The QML method has also found application in the estimation of the power spectrum of the CMB intensity and polarization \citep{Tegmark1,tegmark_pol} and has been recently applied to the latest releases of WMAP data \citep{gruppusowmap5,paci,gruppusowmap7}.
A QML estimator was already used to measure the CMB-LSS cross-correlation only by \cite{padmanabhan}, 
however our implementation is different in few important aspects: the inversion 
of the matrices is implemented here using the single value decomposition (see also Section 3.1) and 
all the three spectra - $TT, TG, GG$ - are computed for all the multipoles in the range of interest. 

Another purpose of this work is to apply our methodology to available public CMB and LSS data, namely 
WMAP 7 year {\citep{jarosik}} and NRAO VLA Sky Survey (NVSS) data \citep{condon}. 
NVSS has been one of the most widely used surveys in the context of ISW studies because 
the radio galaxies it surveys are at high redshifts and it covers a large sky fraction of the sky; 
however, contradicting claims about the evidence of its non-vanishing correlation with CMB exist in 
the literature \citep{PBM,sawangwit} (see also \cite{dupe} for an exhaustive 
compilation of existing results). It is therefore important to apply an optimal methodology to address and quantify the evidence of 
cross-correlation between the most recent large scale CMB measurement and one of the largest LSS survey available.

Our paper is organized as follows: in Section \ref{sect:method} we describe the QML method and give technical details of our implementation of it; Section \ref{sect:montecarlo} discusses our tests of the implementation in simulated maps. 
In Section \ref{sect:data} we report the APS estimates obtained from 
{WMAP 7 year} and {NVSS} data, and then we use these estimates of the cross-correlation
in Section \ref{sect:constrain} to derive constraints on the present critical density due to the cosmological constant. 
Finally in Section \ref{sect:concl} we draw our conclusions.

\section{Methodology}
\label{sect:method}

\subsection{The QML approach} 

The quadratic maximum likelihood method for power spectrum estimate of CMB anisotropies was introduced 
by \cite{Tegmark1} and later extended to polarization by \cite{tegmark_pol}.
Previously, a QML was employed to measure the cross-correlation between CMB and LSS only by \cite{padmanabhan} 
(see also \cite{Ho}). The code in \cite{padmanabhan} estimated only the cross-correlation power 
spectrum only, with a fast and approximated algorithm to invert matrices and used the approximation of a block 
diagonal covariance matrix. In what follows we shall describe the QML method for the whole CMB-LSS data and our 
implementation which does not depend on the simplifying assumptions used in \cite{padmanabhan}. 


Given a CMB map in temperature and a galaxy survey ${\bf x=(T,G)}$ (vector in pixel space), the QML provides an estimator of the angular 
power spectrum 
$\hat {C}_\ell^X$ - with $X$ being one of $TT, TG, GG$.  This estimator is given by    
\be
\label{eq:estim}
\hat{C}_\ell^X = \sum_{\ell' X'} (F^{-1})^{X X'}_{\ell\ell'} \left[ {\bf x}^t
{\bf E}_{\ell'}^{X'} {\bf x}-tr({\bf N}{\bf E}_{\ell'}^{X'}) \right],
\ee
where the $F^{X X'}_{\ell \ell '}$ is the Fisher matrix defined as
\be
\label{eq:fisher}
F_{\ell\ell'}^{X X'}=\frac{1}{2}tr\Big[{\bf C}^{-1}\frac{\partial {\bf C}}{\partial
  C_\ell^X}{\bf C}^{-1}\frac{\partial {\bf C}}{\partial C_{\ell'}^{X'}}\Big],
\ee
and the $E$ matrix is given by
\be
\label{eq:Elle}
{\bf E}_\ell^X=\frac{1}{2}{\bf C}^{-1}\frac{\partial {\bf C}}{\partial
  C_\ell^X}{\bf C}^{-1}.
\ee
${\bf C} ={\bf S}(C_{\ell}^{X})+{\bf N}$ being the total global covariance matrix including the signal $\bf S$ and noise $\bf N$ contributions.  
$C_\ell^X$ is called the fiducial theoretical power spectrum and also is used to create the simulated maps used to test the method in Sec. \ref{sect:montecarlo}. 

Although an initial assumption is needed for this fiducial power spectrum, 
the QML method provides unbiased estimates of the power spectrum of the map regardless of this initial guess 
\be
\langle\hat{C}_\ell^X\rangle=C_\ell^{X} . 
\label{unbiased}
\ee
Here the average is taken over the ensemble of realizations based on the input spectrum $C_\ell^X$.  (See Sec. \ref{sect:montecarlo} for more details.)
The assumed fiducial power spectrum can impact the error estimates, but in practice we start near enough to the true result to be able to neglect this effect.  
The QML method is also optimal, since it can provide the smallest error bars allowed by the Fisher-Cramer-Rao inequality,
\be
\langle\Delta\hat{C}_\ell^X
\Delta\hat{C}_{\ell'}^{X'} \rangle= ( F^{-1})^{X \, X'}_{\ell\ell'}, 
\label{minimum}
\ee
where 
\be
\Delta\hat{C}_{\ell'}^{X'} = \hat{C}_\ell^X - \langle\hat{C}_\ell^X\rangle,
\ee
and the averages, as above, are over an ensemble of realizations.

Our implementation of the QML method is fully parallelized (MPI) and written in Fortran 90.
The inversion of the covariance matrix ${\bf C}$ scales as ${\cal O} (N_{\rm pix}^3)$.
The number of operations is roughly driven, once the inversion of the total covariance
matrix is done, by the matrix-matrix multiplications to build the operators ${\bf E}^{X}_{\ell}$ in
Eq. (\ref{eq:Elle}) and by calculating
the Fisher matrix $F^{X X'}_{\ell \ell'}$ given in Eq. (\ref{eq:fisher}).
The number of operations that are needed to build these matrices scales as $O(N^{7/2}_{\rm pix})$.
This scaling makes clear that the QML method can treat only a limited number of pixels. Therefore 
in the context of an all sky observations it can be applied only at modest resolution.


\subsection{Fiducial spectra}
\label{fiduspe}

For our fiducial model, we assume the concordance $\Lambda$CDM model, with parameters derived from the 
WMAP7 best fit.  Thoughout this work, we assume an equation of state of the dark energy fixed at $w =-1$.
With these assumptions, it is straight forward to calculate the expected power spectra $C_{\ell}^{GG}$ and $C_{\ell}^{TG}$ :
\begin{equation}
C_{\ell}^{GG} = 4 \pi \int \frac{dk}{k} \Delta^2 (k) I_\ell^{G^{2}} (k) \,,
\end{equation}
\begin{equation}
C_{\ell}^{TG} = 4 \pi \int \frac{dk}{k} \Delta^2 (k) I_\ell^{ISW} (k) I_\ell^{G} (k) \,,
\end{equation}
respectively. $\Delta^2 (k)$ is the {logarithmic matter primordial power spectrum, and the filters of the
galaxy density distribution ($I_\ell^{G}$) and the ISW ($I_\ell^{ISW}$) are given by:}
\begin{equation}
I_\ell^{G} (k) = \int dz \, b(z) \, \frac{d N}{d z} \delta_M (k,z) j_\ell \left( k \chi(z) \right) \,,
\end{equation}
\begin{equation}
I_\ell^{ISW} (k) = -2 \int dz \, e^{-\tau} \, \frac{d \Phi}{d z} j_\ell \left( k \chi(z) \right) \,.
\end{equation}
Here, $\frac{d N}{d z}$ is the redshift distribution of the galaxy survey in question, and we have implicitly used the fact that the density contrast in the galaxy survey tracks the matter density contrast as:
\begin{equation}
\delta_G ( \hat{n} )= \int dz \, b(z) \, \frac{d N}{d z} \, \delta_M (\hat{n} ,z) \,.
\end{equation}
It is well known that the late ISW-LSS cross-correlation depends not only on the matter fluctuations on large scales, but also on 
how these are related to the observed galaxy distribution, determined by the the product $b(z) dN/dz$.  This can be simultaneously estimated using the measurement of $C_\ell^{GG}$, also exploiting the QML method.

\subsection{Numerical Improvements}
\label{blockinv}

For the reasons discussed above, the QML method is quite computationally expensive and prohibitive at high resolution. 
We discuss here some changes which can improve the numerics  and decrease substantially the execution time  with a negligible loss of accuracy.

The predicted  $C_{\ell}^{TG}$ is generally non-zero, and its measurement is the primary object of the ISW measurements. 
However, it is expected to be relatively small, even for the largest scales, so it is a good approximation to assume $C_{\ell}^{TG} = 0$ for the fiducial model, which is used to build the covariance matrix.   Further, the noise matrix $\bf N$ may be assumed to be uncorrelated between the CMB and the galaxy measurements.  
Under these assumptions, the Fisher matrix becomes block diagonal and the three spectra $\hat{C}_{\ell}^{TT}$, $\hat{C}_{\ell}^{TG}$, 
$\hat{C}_{\ell}^{GG}$ can be estimated independently from each other. This reduces the computation cost of the Fisher 
matrix by $50 \%$ with respect to the problem with the full covariance. Moreover estimating just $\hat{C}_{\ell}^{TG}$ 
the computational cost of the problem decreases by a further factor of $1/6$, as in \cite{padmanabhan}.

In order to apply the algebra of the QML method, described in 
Eqs. (\ref{eq:estim}-\ref{eq:Elle}), {one must} 
build the covariance matrix ${\bf C}$ in pixel space
and the Fisher matrix ${\bf F}$ in $\ell$ space. The latter is the most expensive task at computationally, largely because it requires the inversion of the pixel space covariance matrix ${\bf C}$.
This inversion can also introduce numerical errors since its eigenvalues naively span
several orders of magnitude.

To bypass this issue, we have used inversion-routines only on numerically homogeneous blocks thanks to the 
following expressions. Given a general matrix $A$ in block form, 
\begin{equation}
   A = \left (
   \begin{array}{cc}
    A_{11} & A_{12} \\
    A_{21} & A_{22}
   \end{array}
   \right),
   \end{equation}
where $A_{11}$ and $A_{22}$ are non-singular square matrices, then
it can be shown that the inverse of $A$ is
\begin{equation}
A^{-1} = \left (
   \begin{array}{cc}
    B_{11} & - B_{11} A_{12} A_{22}^{-1} \\
    - A_{22}^{-1} A_{21} B_{11} & A_{22}^{-1} + A_{22}^{-1} A_{21} B_{11} A_{12} A_{22}^{-1}
   \end{array}
   \right),
\end{equation}
with
\begin{equation}
B_{11} = (A_{11} - A_{12} A_{22}^{-1} A_{21})^{-1} \,.
\end{equation}

For our purposes, we partition the TT, TG and GG blocks of ${\bf C}$, so that $A_{11}$ is the covariance related to the CMB temperature
sector and $A_{22}$ relates to the covariance of the galaxy sector.   Thus, assuming a fiducial model without any cross-covariance simplifies the inversion calculation significantly. 
This technique is also applied to the Fisher matrix inversion in multipole space 
(with $A_{11} = F^{T T}_{\ell \ell'}$), obtaining a much better precision with respect to the brute force inversion. 
 

\section{Validation with simulated maps}
\label{sect:montecarlo}

In order to test our implementation of the QML method, we created simulated galaxy count maps and CMB temperature 
anisotropies following the recipe described in \cite{Boughn:1997vs}  (see also \cite{Barreiroetal2008} and \cite{Giannantonioetal2008}). 
We employ the HEALPix \footnote{http://healpix.jpl.nasa.gov/} program \textit{synfast} {\citep{gorski}} which allows one to create  
$a_{\ell m}$ such that
\be
\langle a^{Y}_{\ell m} {a^{Y^{\prime}}_{\ell^{\prime} m^{\prime}}}^{\star} \rangle = 
{C}^{Y Y^{\prime}}_{\ell}
\delta_{\ell \ell^{\prime}} \delta_{m m^{\prime}},
\ee
where $Y,Y'=T,G$. The total map for the CMB anisotropies $a^T_{\ell m}$ is simulated 
as the sum of three different maps
\be
a^T_{\ell m} = a^{\rm{ISWc}}_{\ell m} + a^{\rm{ISWu}}_{\ell m} + a^{\rm{prim}}_{\ell m}, 
\ee
where $a^{\rm{ISWc}}_{\ell m}$ represents the fully correlated ISW effect with the galaxy distribution,
$a^{\rm{ISWu}}_{\ell m}$ is the uncorrelated part of the ISW effect and $a^{\rm{prim}}_{\ell m}$ is the 
primordial CMB signal. These amplitudes are given by
\be
a^{\rm{ISWc}}_{\ell m}  =  \xi_a \frac{{C}^{\rm{TG}}_\ell}{\sqrt{{C}^{\rm{GG}}_\ell}},
\ee
\be
a^{\rm{ISWu}}_{\ell m}  = \xi_b \sqrt{{C}^{\rm{ISW}}_\ell - \frac{({C}^{\rm{TG}}_\ell)^2} {{C}^{\rm{GG}}_\ell}},
\ee
\be
a^{\rm{prim}}_{\ell m}  \ \ =  \xi_c \sqrt{{C}^{\rm{TT}}_\ell - {C}^{\rm{ISW}}_\ell}.
\ee
In addition for the galaxy count maps we consider
\be
a^{G}_{\ell m} = \xi_a \sqrt{{\rm{C}}^{GG}_\ell},
\ee
where $\xi$'s are {Gaussianly} distributed complex random numbers,  with zero {mean and unit variance}.
They are the seeds of the simulations and satisfy $\langle \xi_a \xi^{*}_{a^{\prime}} \rangle = \delta_{a a'}$. 
In this way it can be shown that
\be
\langle a^{\rm{T}}_{\ell m} a^{\rm{T} \, *}_{\ell m} \rangle = {C}^{\rm{TT}}_\ell,
\ee
\be
\langle a^{\rm{G}}_{\ell m} a^{\rm{G}\,*}_{\ell m} \rangle = {\rm{C}}^{GG}_\ell .
\ee
\be
\langle a^{\rm{T}}_{\ell m} a^{\rm{G}\,*}_{\ell m} \rangle = {\rm{C}}^{TG}_\ell .
\ee

\begin{figure}
\includegraphics[width=9cm,height=5.5cm,angle=0]{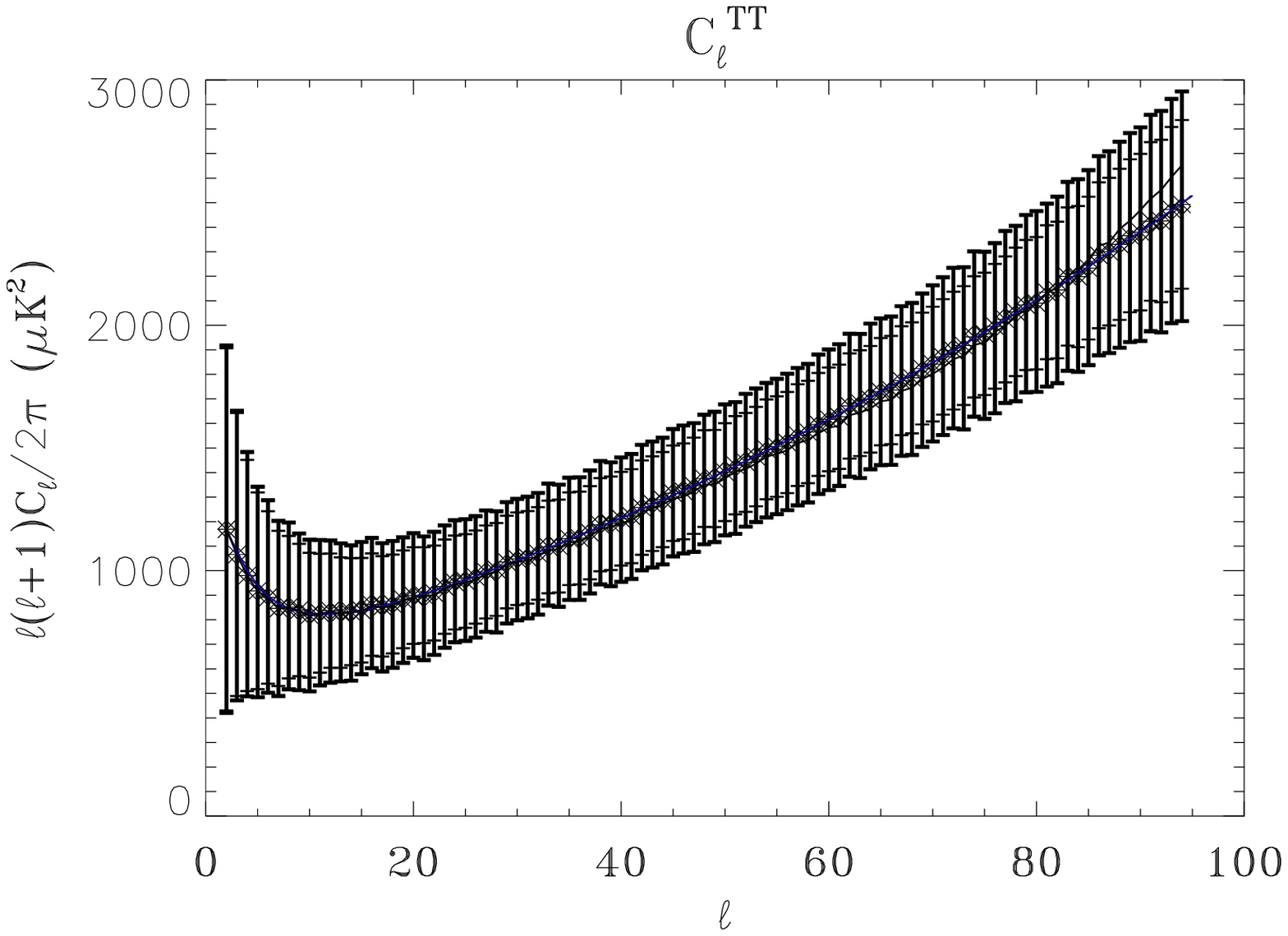}
\includegraphics[width=9cm,height=5.5cm,angle=0]{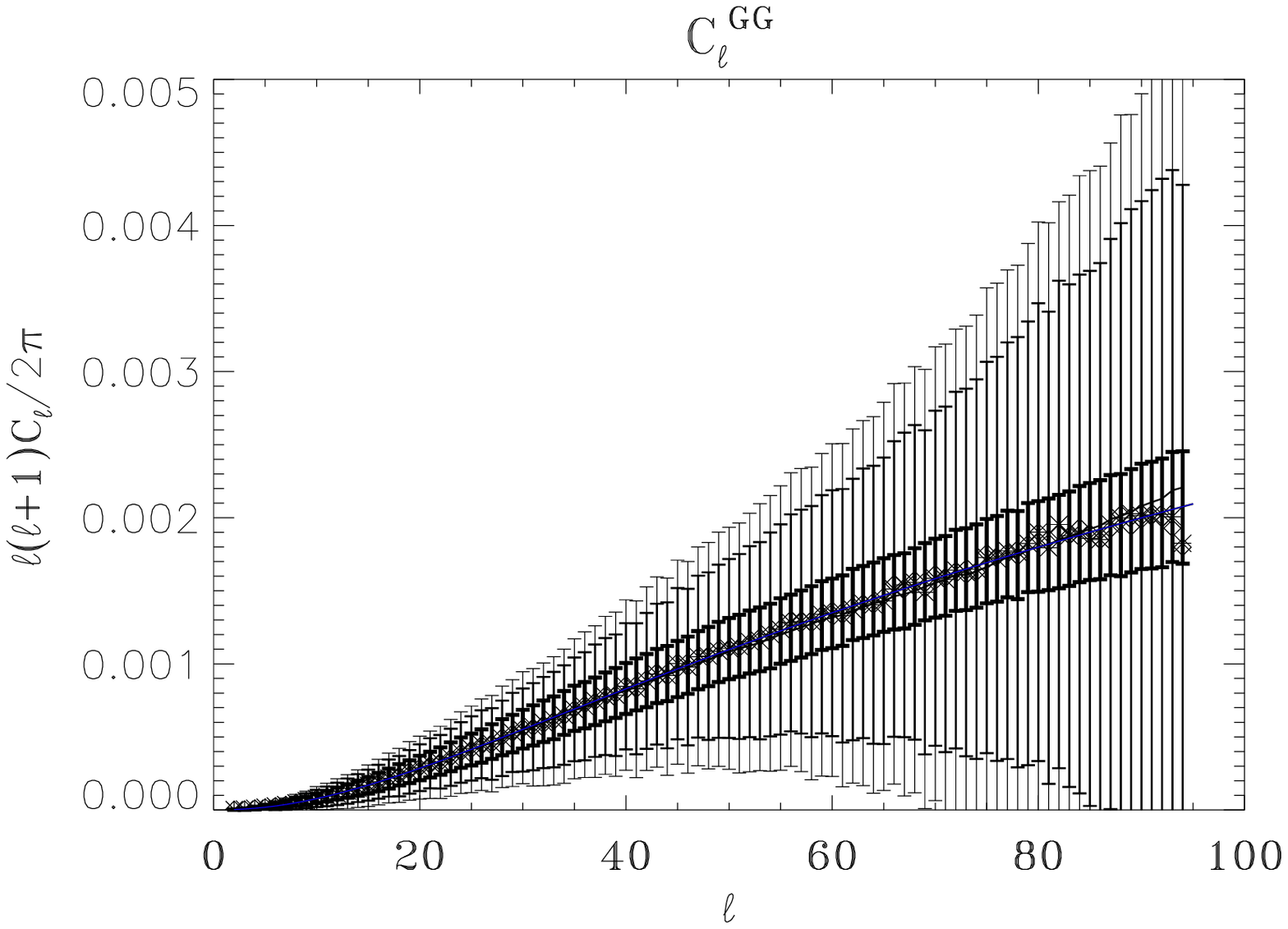}
\includegraphics[width=9cm,height=5.5cm,angle=0]{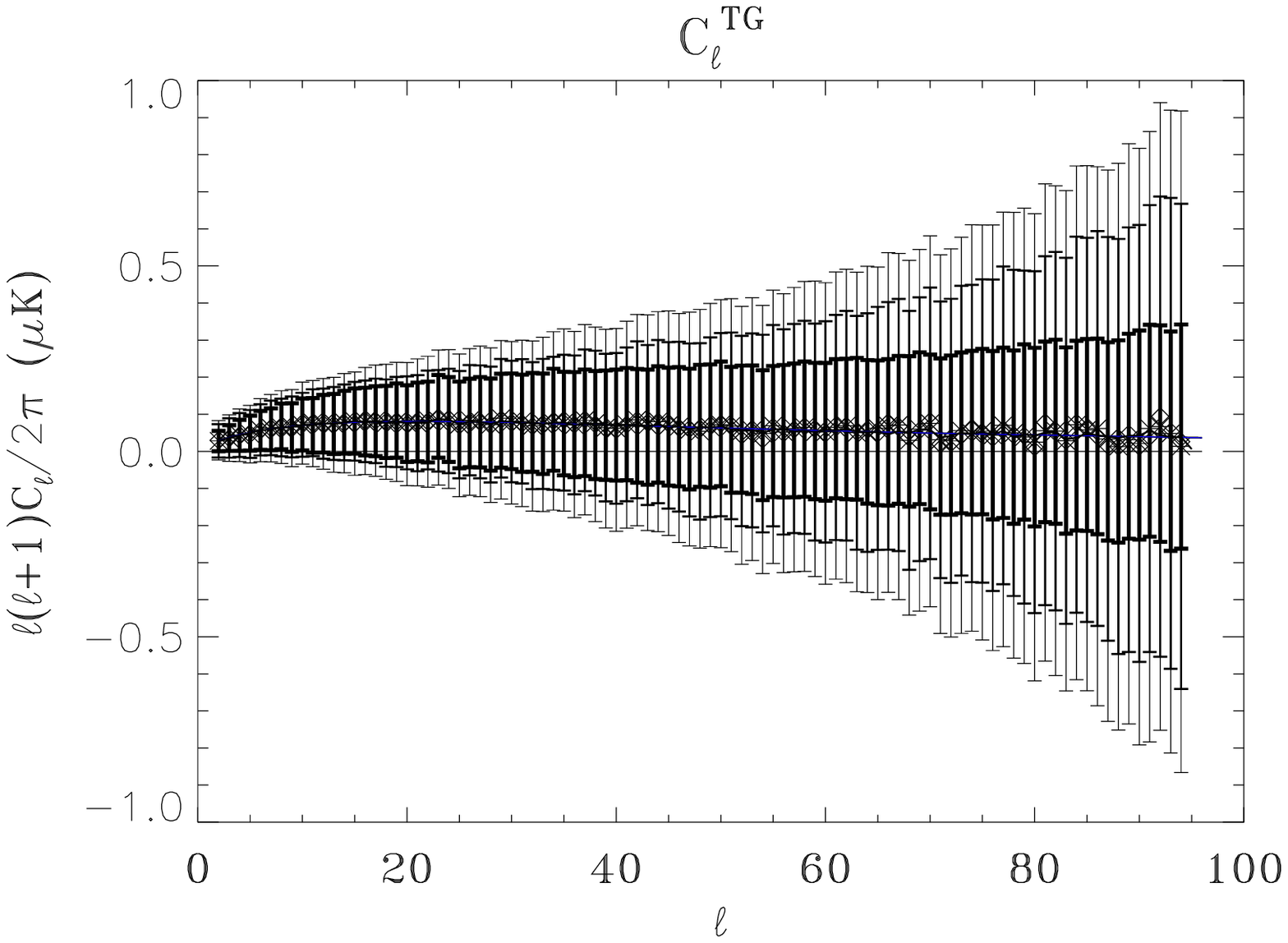}
\caption{The average estimates for the Monte Carlo validation: the upper and middle panels show the TT and GG auto-spectra, respectively, and the lower one is the TG cross-spectrum. 
We compare results for three cases: using realistically masked maps without noise in the LSS maps (thick error bars), using full sky maps with NVSS-like shot noise (solid line error bars), and assuming both masked maps and NVSS-like shot noise (light dark error bars). 
We can see that average power spectra from the QML all agree very well with the underlying fiducial theoretical 
power spectra (blue lines). The error bars change according to the noise level in the LSS map and the fraction of the sky considered.
The dark lines are the average of 
the \textit{anafast} estimates, which are slightly biased at high $\ell$ in the two auto-spectra.} 
\label{fig:Montecarlo}
\end{figure}

\begin{figure}[h]
\includegraphics[width=9cm,height=5.5cm,angle=0]{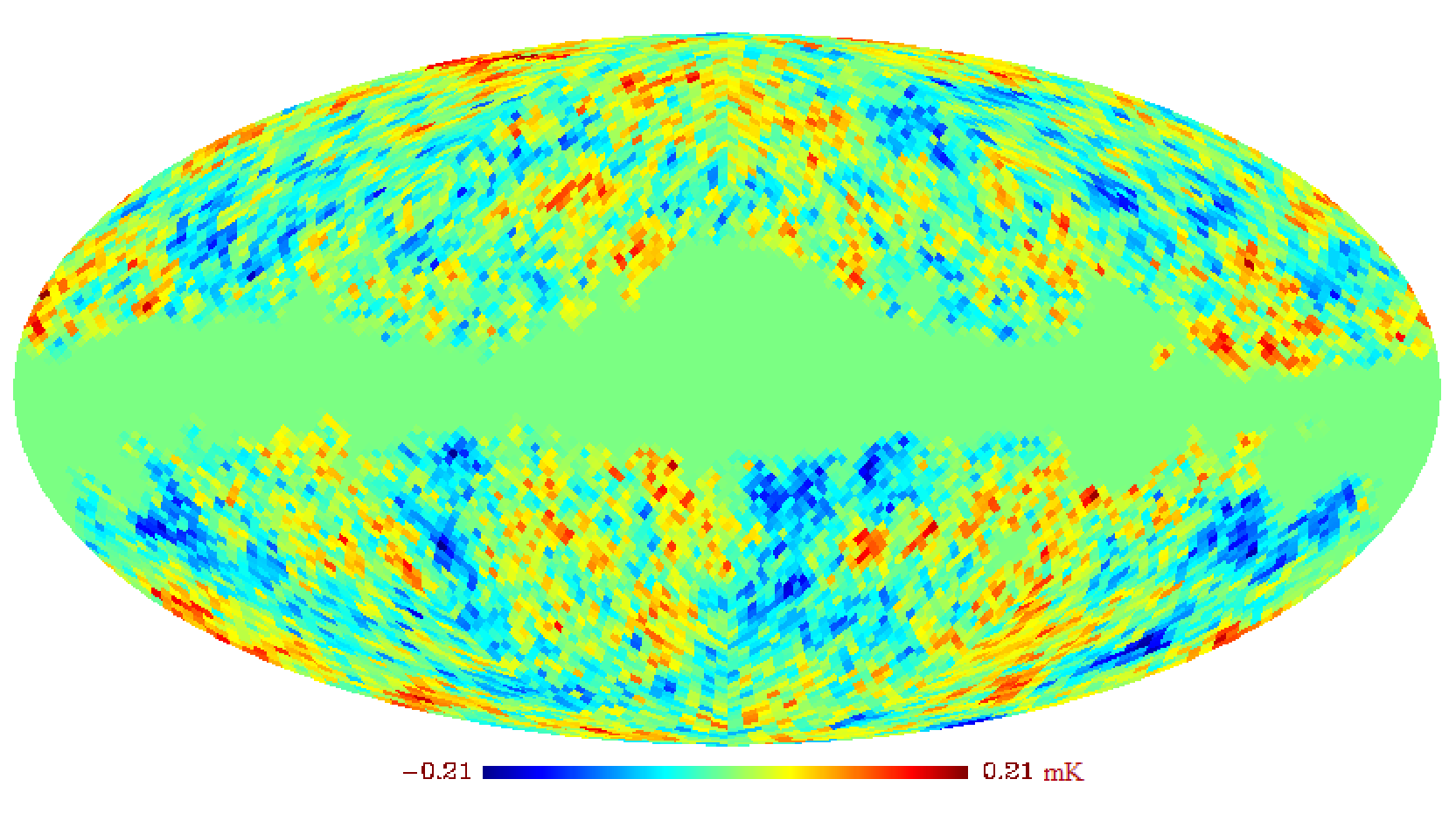}
\includegraphics[width=9cm,height=5.5cm,angle=0]{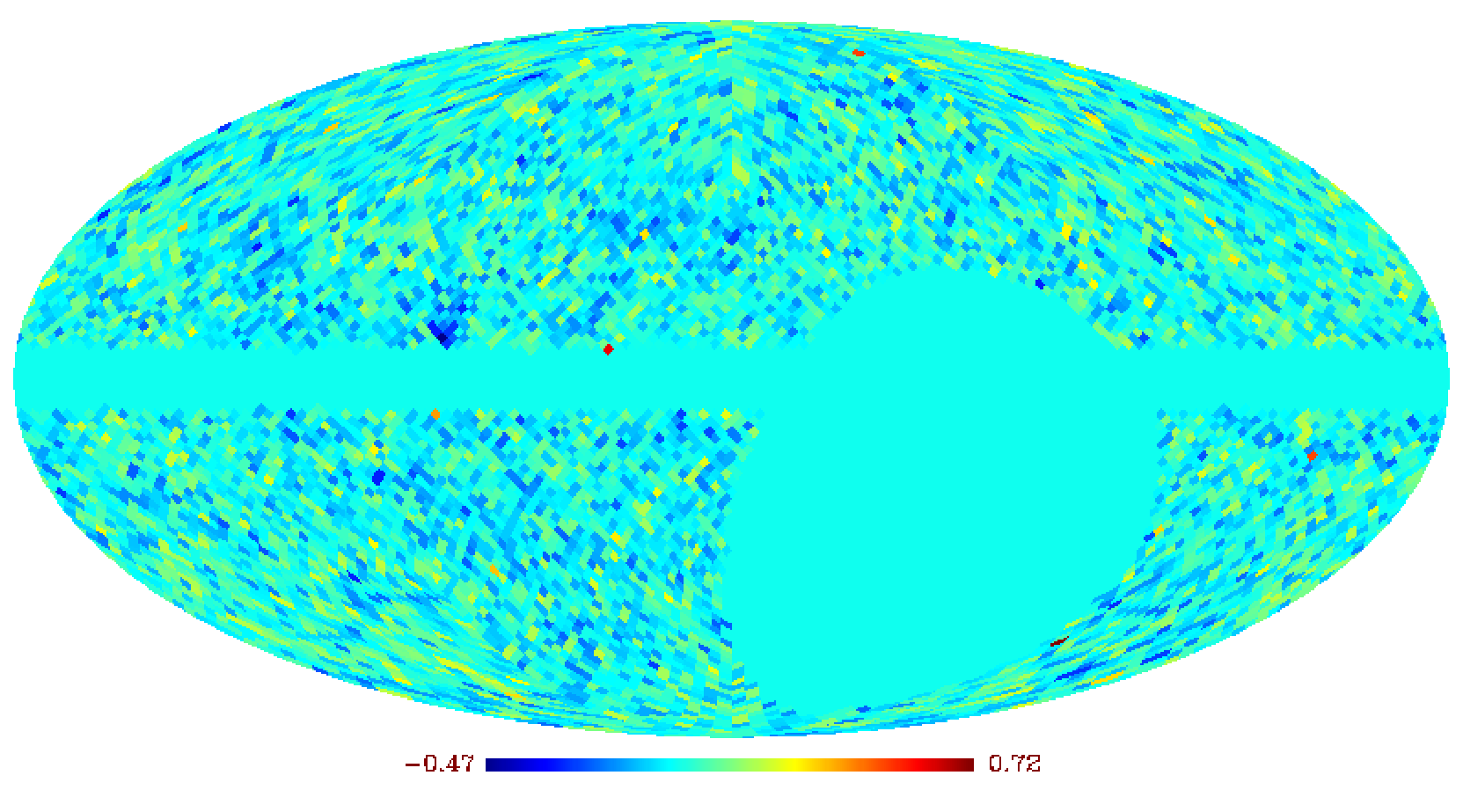}
\caption{WMAP 7 yr (top) and NVSS (bottom) maps at HEALPIX resolution $N_{\rm side} = 32$ used in this analysis, with the respective masks. 
In the displayed NVSS map, the threshold 
flux is $2.5$mJy and the corrections for systematics in declination has been applied.}
\label{fig:maps}
\end{figure}

\begin{figure}
\includegraphics[width=6.5cm,angle=-90]{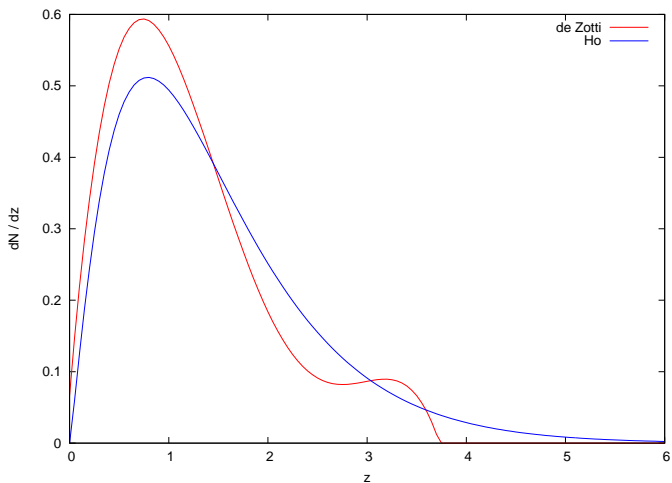}
\caption{Comparison of the source redshift distribution used in Ho et al. (2008) (blue) and the one with a redshift dependent bias with
assuming the CENSORS source distribution (red).}
\label{honzvsdezottinz}
\end{figure}

\begin{figure*}
\includegraphics[width=8.5cm,height=5.5cm,angle=0]{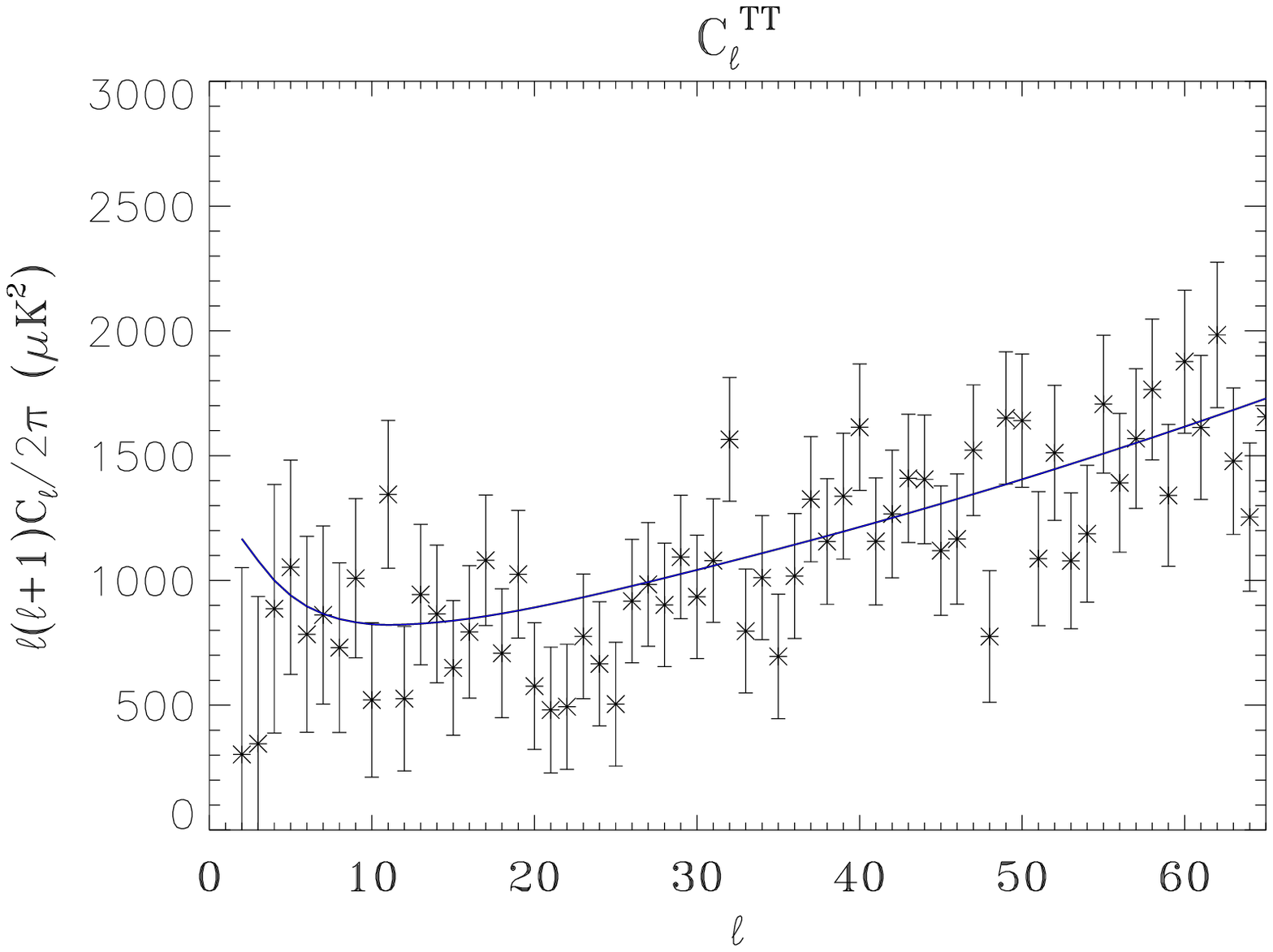}
\includegraphics[width=8.5cm,height=5.5cm,angle=0]{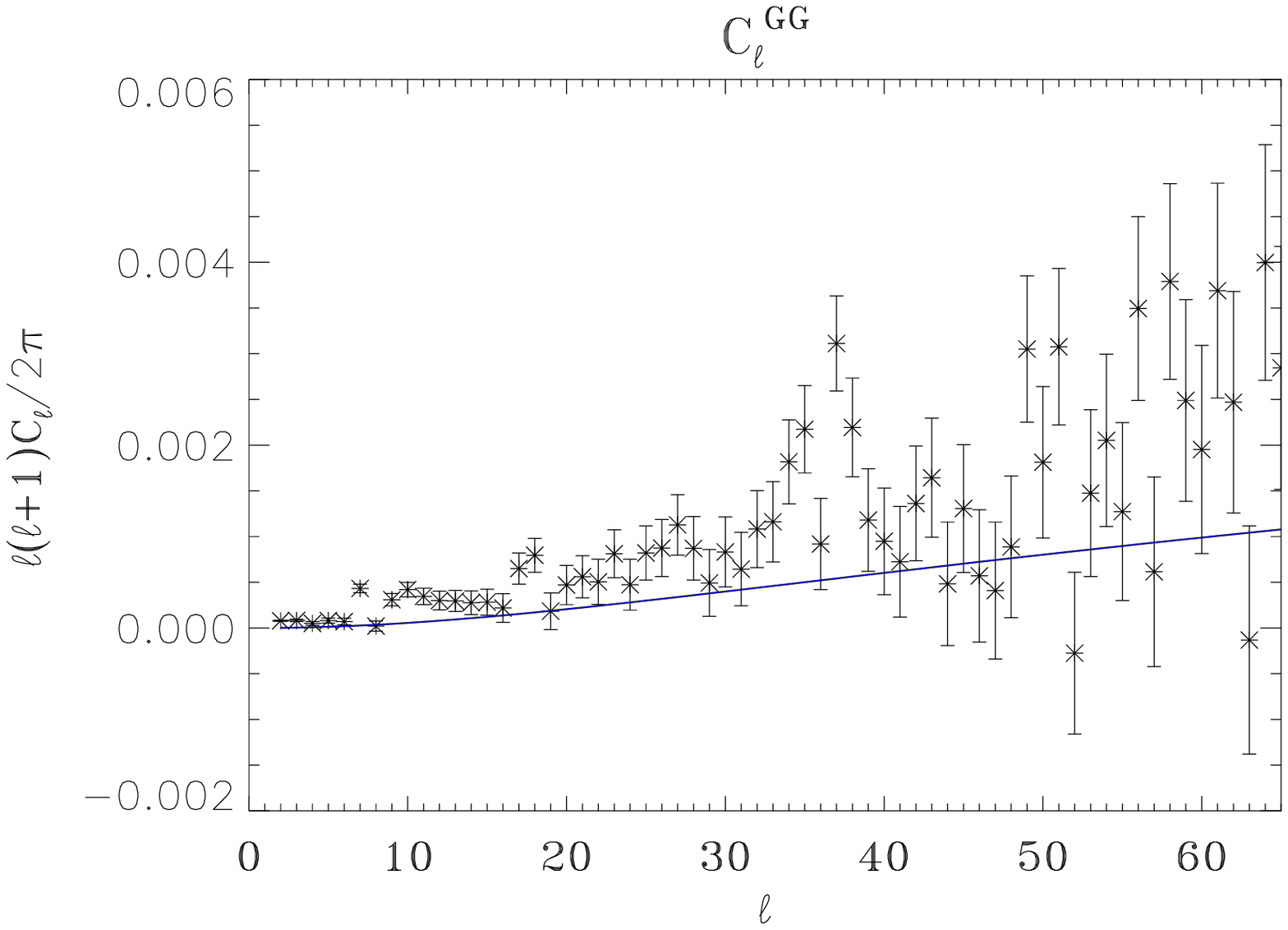} \\
\includegraphics[width=8.5cm,height=5.5cm,angle=0]{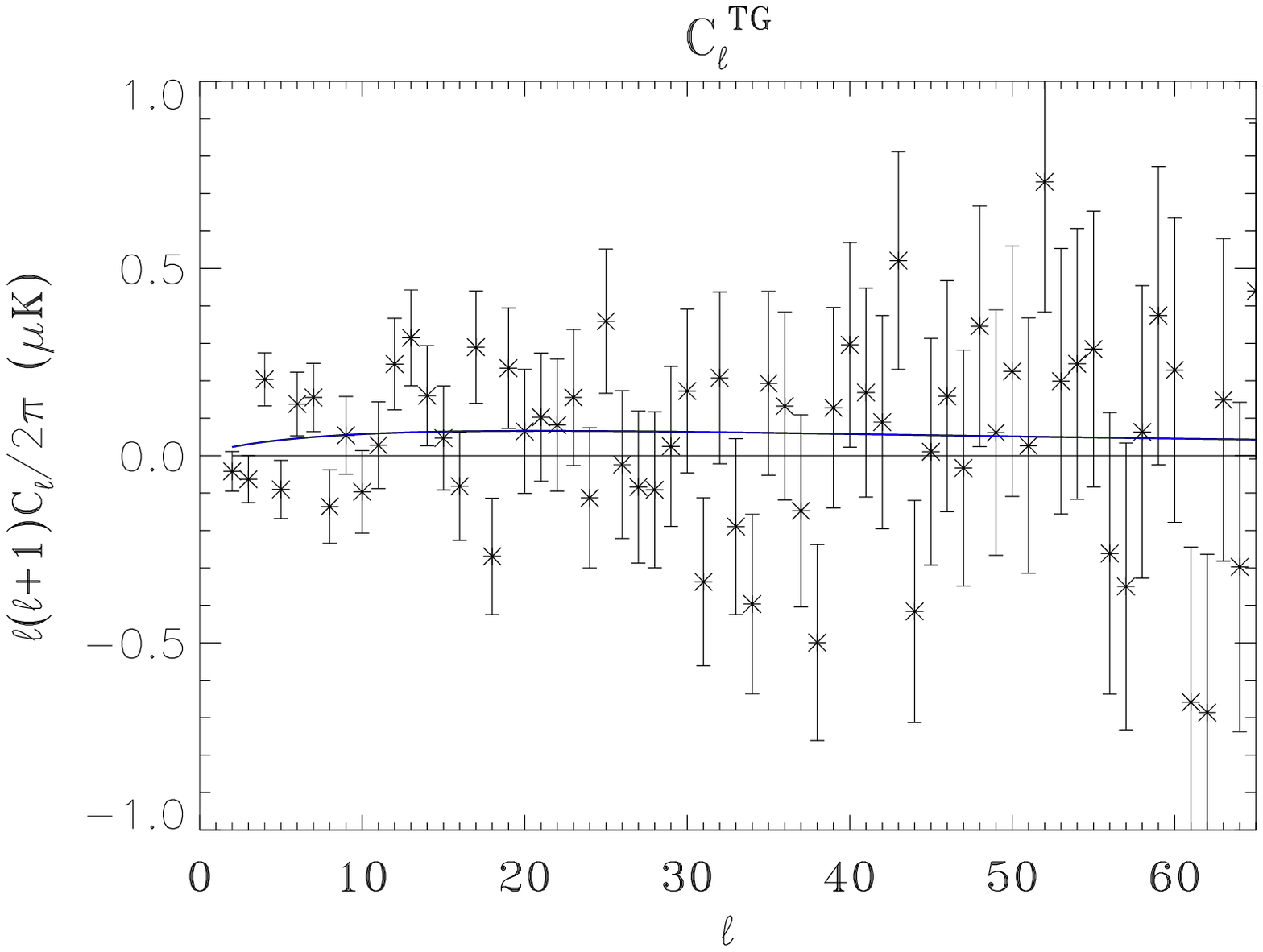}
\includegraphics[width=8.5cm,height=5.5cm,angle=0]{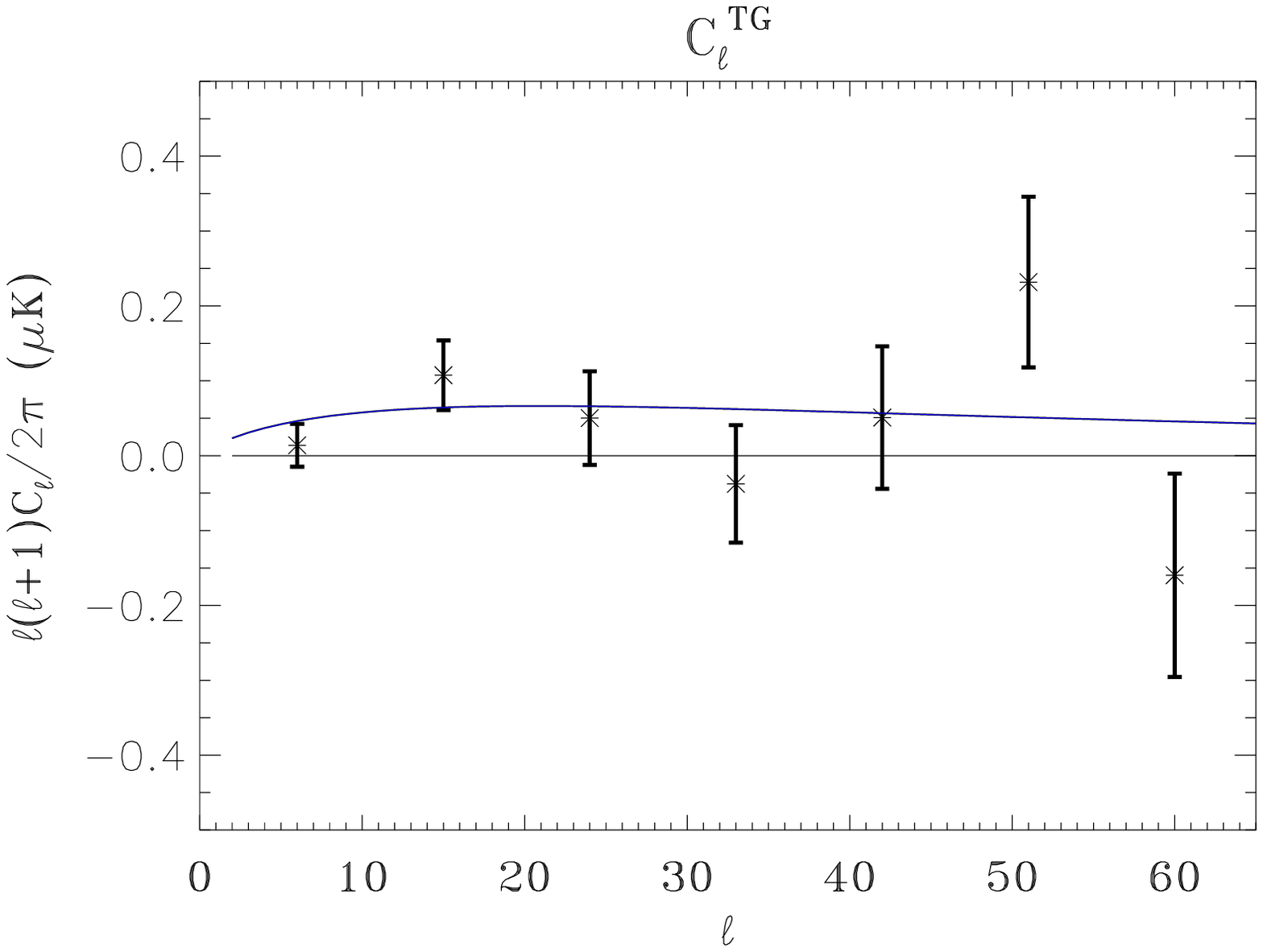}
\caption{APS estimates from WMAP 7 year and NVSS data for the $2.5 mJ$ flux cut. 
The upper panels show the auto-spectra, 
TT and GG, the lower are the cross-spectra, TG, in the no binning case (left panel) and in the binning case (right panel). 
The blue solid lines are always the fiducial power spectra. The TG cross-spectrum in the lower right panel is binned with $\Delta \ell=9$ to 
make clearer the QML estimates respect to the fiducial power spectrum (blue solid line) and the $C_{\ell}^{TG} =0$ (black solid line). 
We can note the QML estimates for the GG auto-spectrum are slightly 
larger than the expected fiducial power spectrum.}
\label{spectraWMAPplusNVSS}
\end{figure*}

{We have tested the QML approach using these Monte Carlo simulations. In particular, we have performed 
$3000$ realizations for CMB and LSS correlated maps at the HEALPix resolution of 
$N_{\rm side} = 32$ \footnote{The number of pixels $N_{\rm pix}$ 
is related to the parameter $N_{\rm side}$ through $N_{\rm pix}=12 N_{\rm side}^2$.}. 
For the multipoles, we consider the range $\Delta \ell = [2,95]$; i.e., up to the Nyquist frequency $3 N_{\rm side} - 1$. 
The standard $\mathrm{\Lambda}$CDM cosmological model~\citep{larson} is assumed, as well a survey characteristics similar 
to the NVSS catalogue~\citep{condon}, namely: a similar sky coverage (see next Section),
a galaxy density number distribution per redshift given by the~\cite{Ho} model, and a bias $b=1.98$.} 

{These simulated maps show that our QML implementation leads to 
unbiased and minimum variance results when considering the realistic case of a masked sky, 
as can be seen by comparing 
the simulations to the projected errors from the Fisher matrix.  
Importantly, we confirm that the method is unbiased and minimum variance
when the signal covariance matrix is block diagonal, i.e. when 
fiducial cross power spectrum $C_{\ell}^{TG}$ is set to zero: with the latter approximation, no 
difference can be appreciated by eyes on the QML estimates and a very small difference can be seen in the likelihood 
constructed by Fisher, which will shown for our application to real maps of WMAP 7 yr and NVSS in Sect. 
}

{It is important to notice that, while on these large-scales the noise contribution  
in WMAP and future ({\sc Planck}) CMB temperature maps is so low that the CMB noise ${\bf N}$ can be safely neglected, this is not necessary true for large scale structure surveys. Depending on the number of sources used as large scale tracers,
the galaxy density map could be significantly affected by Poissonian noise, which must be taken into account.}

{The results from the Monte Carlo validation are summarized in Fig.~\ref{fig:Montecarlo}: the upper, middle and lower panels
show respectively the average estimates for the TT, GG and TG spectra derived from the Monte Carlo simulations. 
Three different scenarios are considered, all of which provide unbiased averaged estimates in good agreement with the fiducial model (blue lines), and they differ only in their error bars. 
The first case corresponds to a masked sky (accounting for the NVSS sky coverage and the WMAP KQ75 mask
with negligible Poissonian shot-noise contribution to the LSS map (given by the thick error bars); second, a full-sky case with a shot-noise like the that expected in NVSS (see next Section for more details) when only sources above 2.5 mJy 
are taken into account (solid line error bars); and, finally a more realistic situation where both, the incomplete sky and the shot-noise are included in the analysis (light dark error bars). The error bars increase when the noise level in the LSS map rises and when the fraction of the sky considered is reduced, 
the latter falling approximatively with the $\sqrt{f_{\rm sky}^T f_{\rm sky}^G}$, as expected.

For comparison, the plots also
include (dark lines) the average \textit{anafast} estimation for the full-sky case (dark lines), based on the simple HEALPix FFT tool. As it can be seen, the \textit{anafast} estimation is slightly biased at high $\ell$ in the two auto-spectra.}


\section{Application to WMAP 7 year and NRAO VLA Sky Survey data}
\label{sect:data}

\begin{figure}
\includegraphics[width=8.5cm,angle=0]{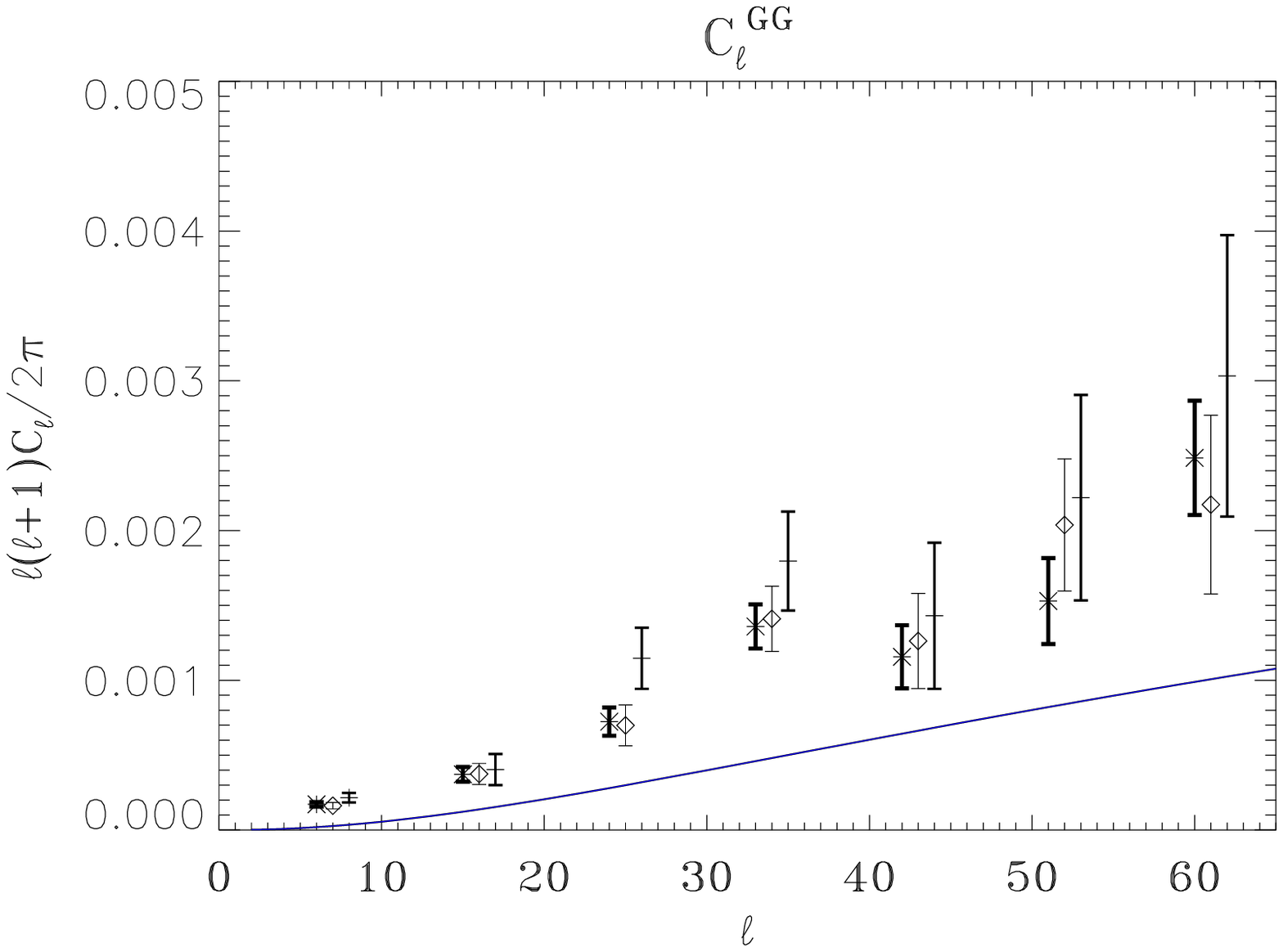} \\
\includegraphics[width=8.5cm,angle=0]{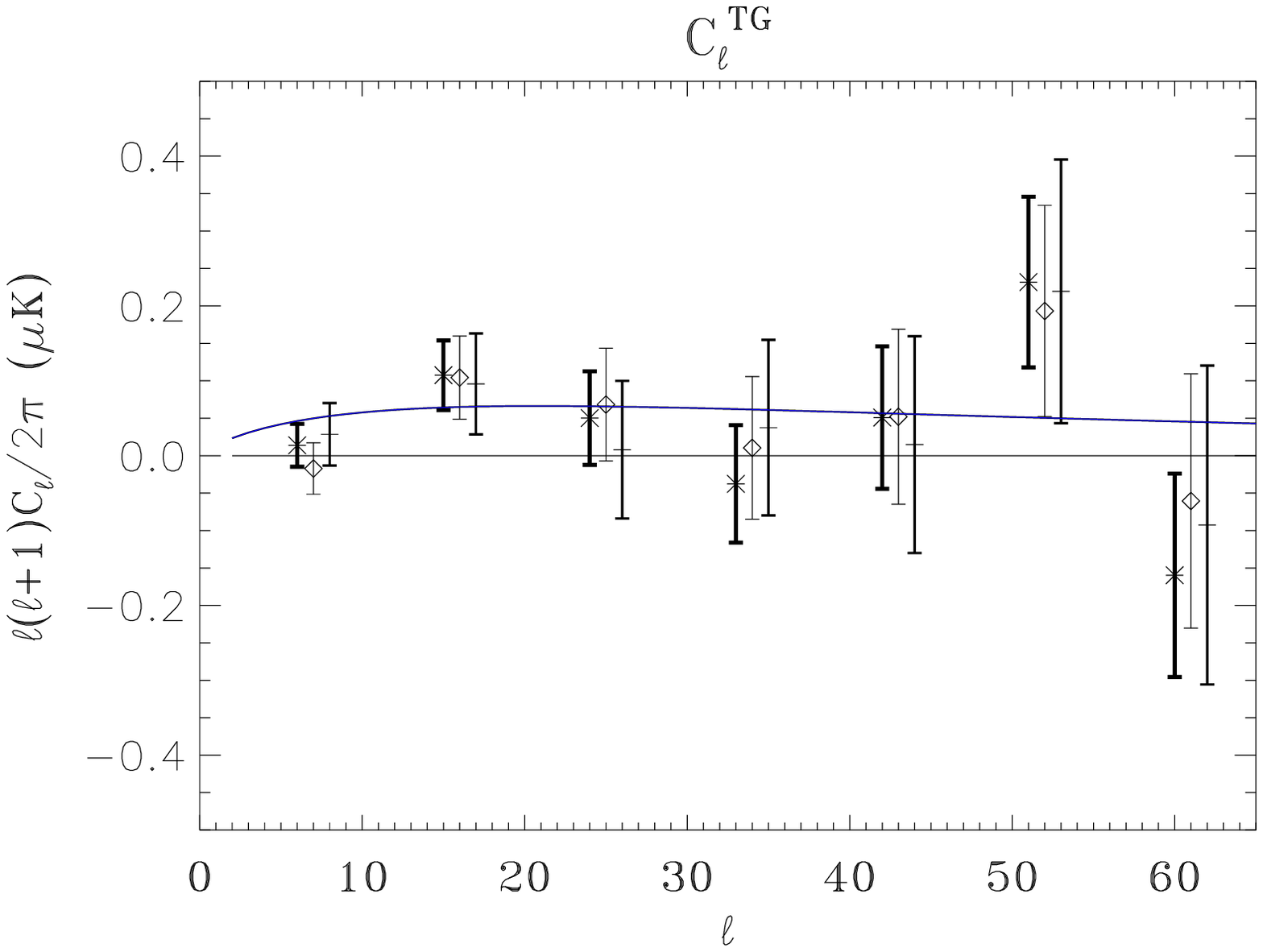}
\caption{Comparison of the binned GG and TG spectra for  
three thresholds in flux, i.e. 2.5 mJ (stars with thick error bars), 5 mJ (diamonds with grey 
error bars) and 10 mJ (cross with error bars).  The results appear largely consistent, while the errors increase as the number of 
sources decrease.}
\label{differentcuts}
\end{figure}

\begin{figure}
\includegraphics[width=8.5cm,angle=0]{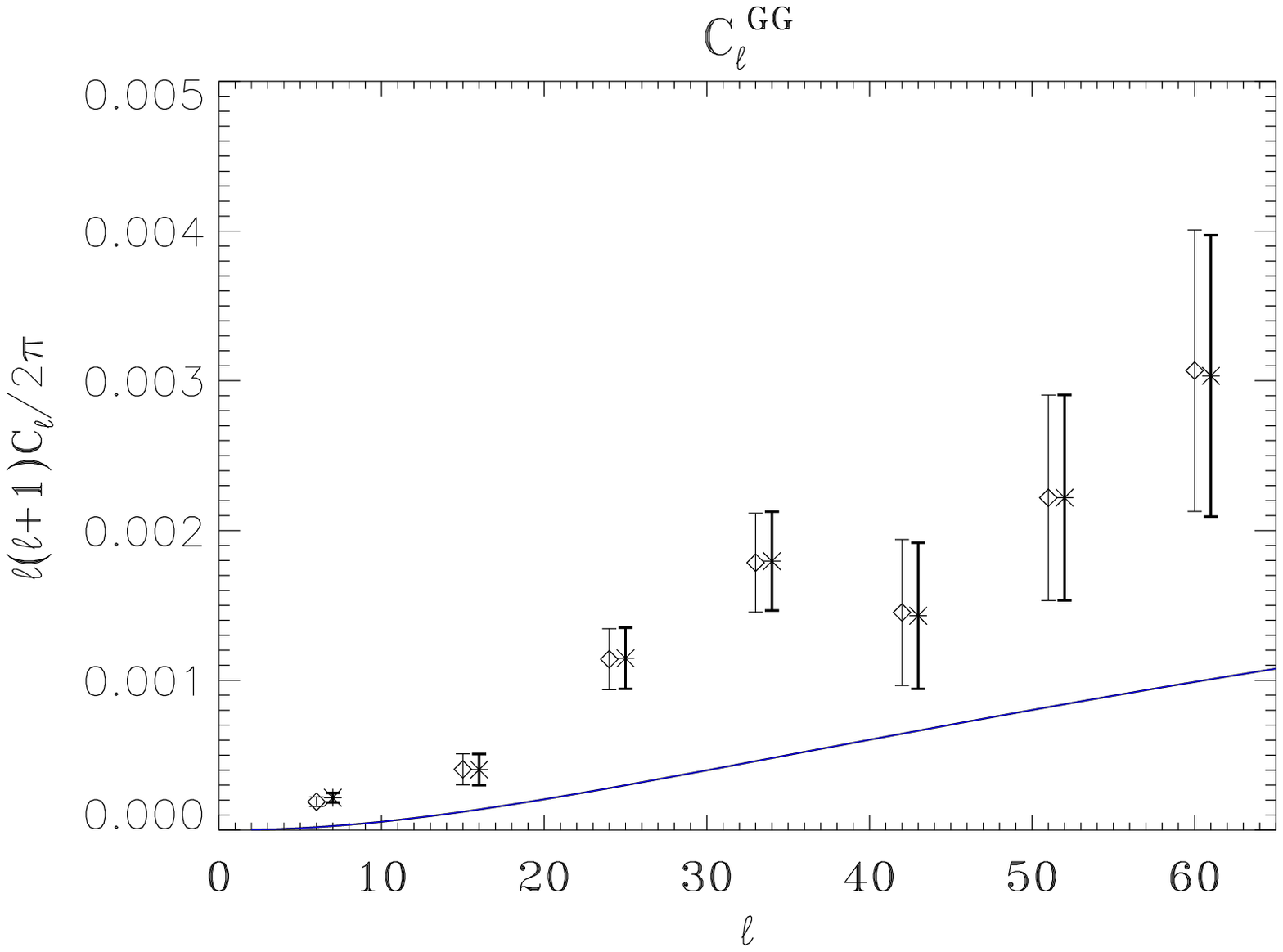} \\
\includegraphics[width=8.5cm,angle=0]{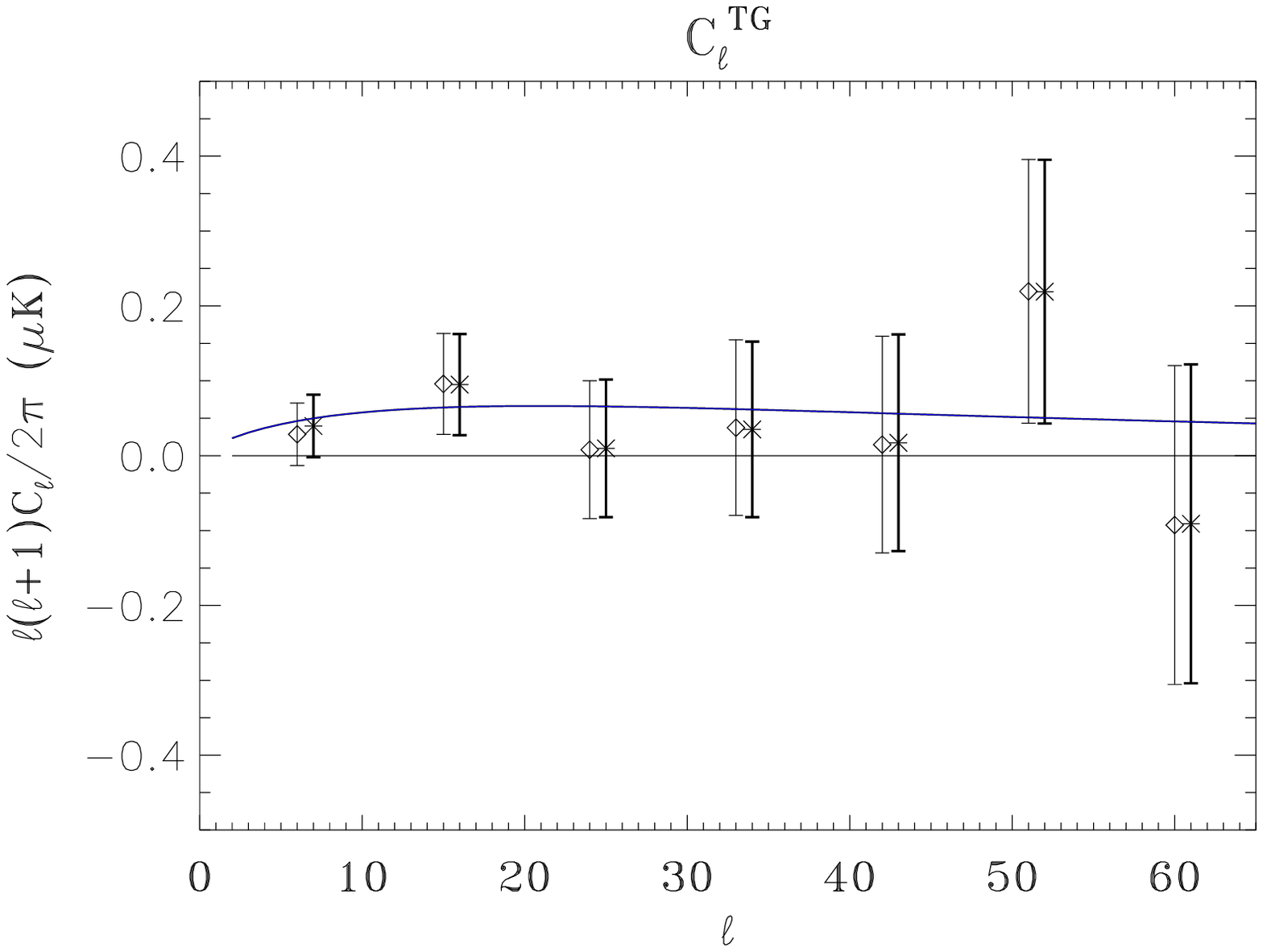}
\caption{Comparison of the binned GG and TG spectra  when the declination correction is included (diamonds)
or not included (cross){, assuming a flux cut of 10mJy}. }
\label{declinationcorrection}
\end{figure}

\begin{figure}
\includegraphics[width=8.5cm,angle=0]{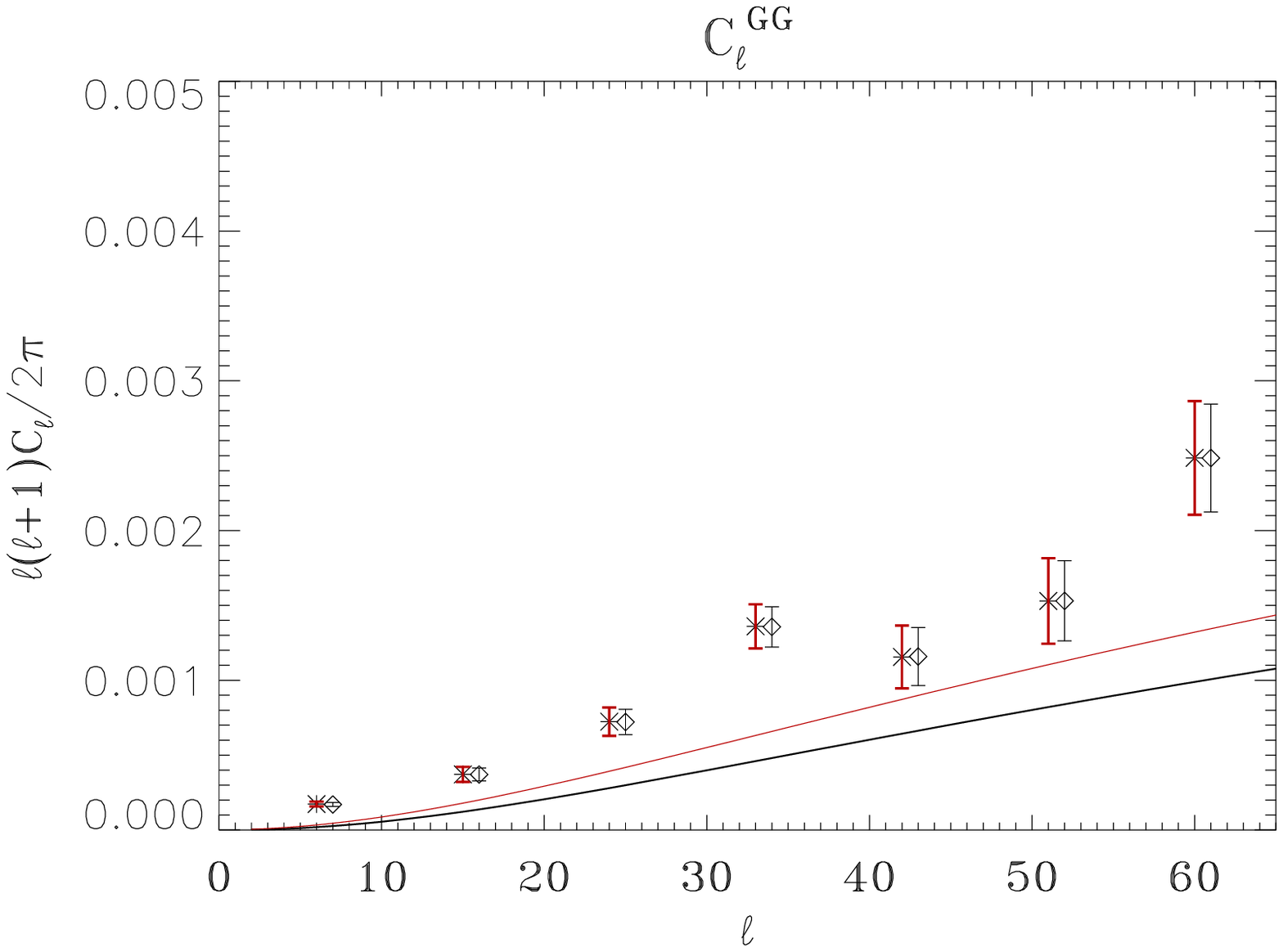} \\
\includegraphics[width=8.5cm,angle=0]{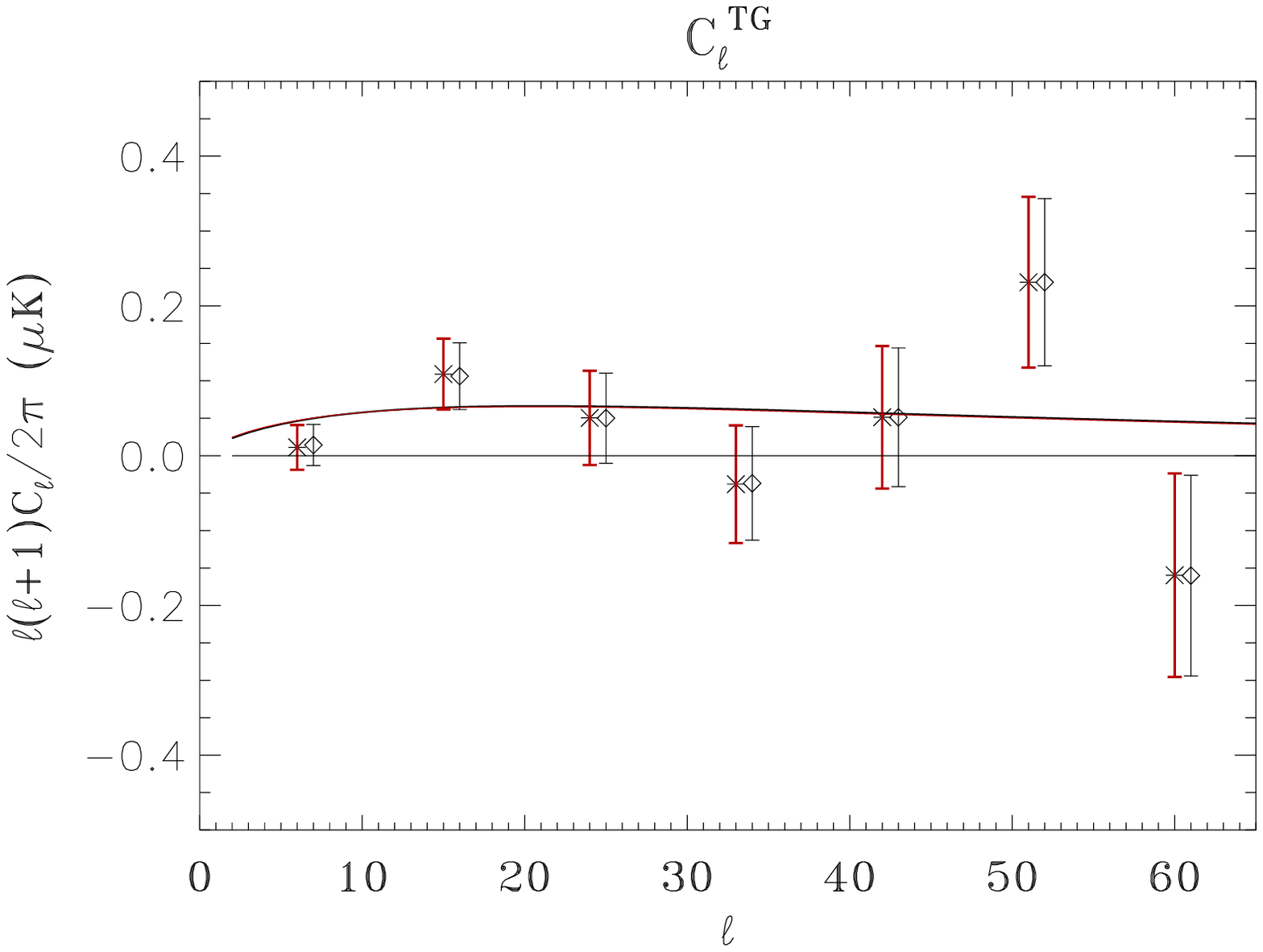}
\caption{Comparison of the binned GG and TG spectra for the fiducial model (black) and the one with a redshift dependent bias with 
the CENSORS source distribution (red). 
Note how the estimates are stable with respect to the change in the fiducial model, and how an evolving bias could alleviate the tension 
between the theoretical GG predictions and the NVSS estomates.}
\label{hovscensors}
\end{figure}

In this Section we describe the application of our QML code to estimating the cross-correlation spectrum between the WMAP 7-year CMB maps and the NRAO VLA Sky Survey (NVSS) data.

\subsection{The maps}

For WMAP data we make use of publicly available products\footnote{http://lambda.gsfc.nasa.gov/}.
In particular, clean maps at the V and W frequency bands have been co-added, using a 
weighting procedure that accounts for the instrumental noise variance per pixel. 
These frequency maps have been cleaned following a template fitting approach~\citep{gold}, 
and are those used by the WMAP tem to perform cosmological tests, such as constraining
non-Gaussianity~\citep{komatsu11}. The co-added map has been degraded from its original $N_{side}=1024$ 
down to $N_{side}=32$, since the angular scales associated to this resolution ($\approx 2^{\circ}$) 
is enough to capture almost all the signal in the CMB-LSS cross-correlation expected from the ISW effect.  Following this, the WMAP 
KQ75 Galactic mask (similarly degraded) is applied to the co-added map, 
in order to mitigate the unavoidable foreground contamination in regions within and near the Galactic plane, 
and also to remove known and intense extragalactic objects such as the Magellenic clouds and large clusters 
near the northern Galactic pole. Finally, the remaining monopole and dipole moments outside the mask have been estimated and removed.   

The NVSS catalogue~\citep{condon} is a radio sample at 1.4 GHz produced with the Very Large Array. 
It covers $\approx 85\%$ of the sky, up to an Equatorial declination of $\delta > -40^\circ$. 
The original survey accounts for $\approx 2\times10^6$ sources with fluxes $>2.5$mJy. This survey has been widely used in the 
context of the ISW studies. It was first used by~\cite{BC} 
to probe the CMB-LSS cross-correlation with the COBE data, and a few years afterwards it was 
successfully used by the same authors with WMAP data, in the first work reporting such cross-correlation~\citep{boughn04}; this was soon followed  
by \citep{noltaetal} with a similar analysis by the WMAP team.  

The survey has a somewhat inhomogenous sensitivity as a function of the equatorial 
declination~\citep[see][for the details]{condon}, resulting in the mean galaxy density that 
artificially varies with the declination. Therefore, some pre-processing is needed in order to mitigate this 
large-scale effect. {One of the procedures used in the literature consists} in defining iso-latitude bands 
(in equatorial coordinates) and imposing that these bands to have the same mean galaxy density. 
In our case, this pre-processing consists of selecting first the sources above a particular flux cut, and then 
defining nine bands of equal area, imposing the same mean galaxy density number for each band.  
Finally, we rotate to Galactic coordinates to compare to WMAP, and then pixelise to a HEALPix resolution of $N_{side} = 32$.

Previous works~\citep[e.g.,][]{noltaetal,vielva} have shown that the particularities of this 
pre-processing do not affect significantly the results and we confirm this below. We also repeat the analysis 
for different thresholds in flux, namely $2.5 \,, 5 \,, 10$mJy, as the higher flux thresholds should be less sensitive to possible declination systematics.

\subsection{Source redshift distribution} 

To interpret the results of our measurements, we must assume some redshift distribution 
$dN/dz$ and potentially redshift dependent bias $b(z)$ for the sample. Given a redshift 
distribution, the average bias can be estimated from the measured QML estimates for 
$C_\ell^{GG}$; however, here we exploit previous measurements for the NVSS sample.

Historically, the redshift distribution was based on models of the sources by 
\cite{dunlop}, and a time-independent bias of 1.6  was derived by \cite{BC}. 
A larger time-independent bias was found by \cite{blakeferreiraborrill}, albeit with a 
different redshit distribution than used by \cite{BC}.
In \cite{Ho}, a new redshift distribution was derived based on a $\Gamma$ 
distribution fit which was constrained to give the cross-correlations measured 
between the NVSS survey and SDSS LRG subsamples: 
\be
\dNdz{\ho} = 
\frac{\alpha^{\alpha}}{z_*^{\alpha+1} \Gamma(\alpha)} z^\alpha {e^{-\alpha z/z_*},}
\label{honz}
\ee
where $z_* = 0.79$ and $\alpha = 1.18$ \footnote{\bf Note that we have corrected 
the normalization factor of the $\Gamma$ distribution assumed in \cite{Ho}.}. 
\cite{Ho} also estimates 
an effective, redshift independent 
value for the bias as $b(z) = 1.98$. 
Finally, we also explore the most recent galaxy 
redshift distribution proposed by \cite{dezottireview}: 
a fourth order polynomial fit to the CENSORS distribution \citep{censors}:
\begin{equation}
\dNdz{\dZ} = 1.29 + 32.37 z -32.89 z^2 +11.13 z^3 -1.25 z^4 \,.
\label{dezottinz}
\end{equation}
A comparison of the two redshift distributions based on Eqs. (\ref{honz}) and (\ref{dezottinz})
is shown in Fig. (\ref{honzvsdezottinz}). For the second redshift distribution in Eq. (\ref{dezottinz}) 
we consider a redshift dependent bias \citep{matarreseetal,moscardinietal}:
\be
b(z) = b_0 + \frac{b_1}{D^\gamma(z)} 
\label{biasz}
\ee
where $D(z)$ is the linear growth factor in 
a $\Lambda$CDM Universe. Following \cite{xia10a}, we choose $b_0 = 1.1$, $b_1=0.6$, $\gamma=1$.
Below we focus on the latter two distributions, and examine how the uncertainties impact 
the derived cosmological constraints.

\subsection{Measurements of the spectra}

In Fig. \ref{spectraWMAPplusNVSS} we present the TT, GG and TG spectra obtained by our QML up to 
$\ell = 64$ ($ = 2 N_{side}$) for the $2.5$ mJ flux cut in NVSS data. 
Since the signal-to-noise for unbinned TG power spectrum is rather poor, we present also the binned power
spectrum $C^{\rm{TG}}_b$ over $\Delta \ell=9$. The binned estimates are simply the average of the unbinned estimates inside the bin.
For plotting purposes, we associate $\sqrt{\sum_{\ell \in \Delta \ell} (F^{-1})_{\ell \ell}^{TG \, TG}/N}$
for the uncertainty in the binned estimate.
Unless otherwise stated, all the maps have been corrected for the declination effect.

To investigate potential systematic problems, we compare the dependence of the TG and GG spectra on the different threshold fluxes for NVSS 
considered here in Fig. \ref{differentcuts} (TT is not shown since it is of course unchanged.) 
Overall the APS estimates agree very well when varying the flux threshold, 
with larger error bars for larger flux threshold, as expected given the fewer objects and resulting larger Poisson errors. 
In Fig. \ref{declinationcorrection} we examine the importance of the correction for declination systematics in NVSS 
for a flux cut of 10 mJ. Our result for GG agrees with \cite{blakeandwall}, confirming that with a conservative flux cut of 10 mJ the declination 
systematics in NVSS is negligible.

\begin{figure}
\includegraphics[width=8cm,angle=0]{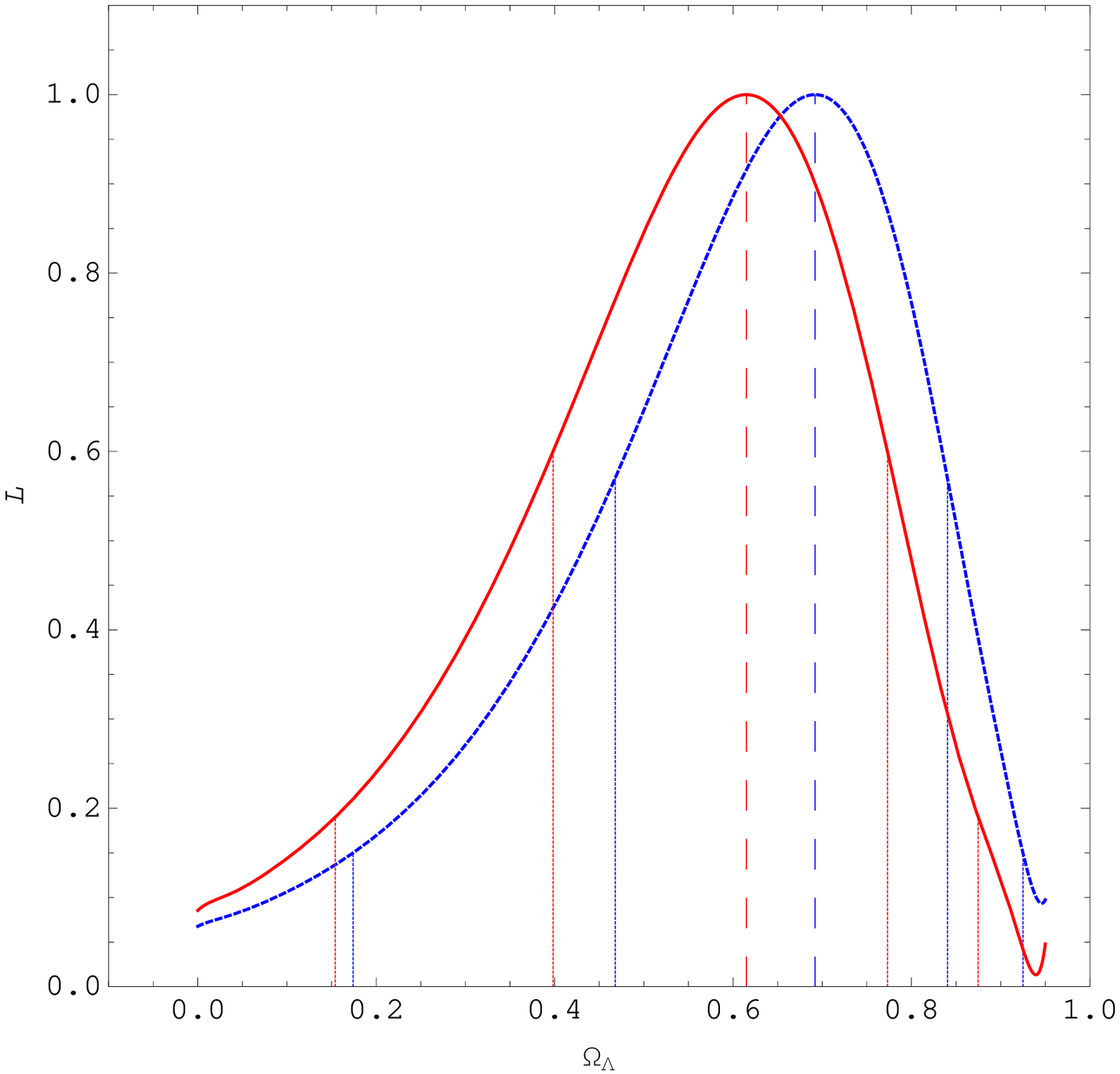}
\caption{Comparison of likelihood contours for ${\Omega}_{\Lambda}$ obtained by the Fisher
prescription in Eq. (\ref{fisher}) for the two choices of redshift distributions: solid for 
Eq. (\ref{honz}) and dashed for Eq. (\ref{dezottinz}).
The threshold flux in NVSS has been chosen as 2.5 mJ.}
\label{likelihood_2p5mJ_deZottiVSHo}
\end{figure}

Finally, in Fig. \ref{likelihood_2p5mJ_deZottiVSHo} 
we compare the GG and TG results obtained by assuming the two redshift distributions 
in Eqs. (\ref{honz}) and (\ref{dezottinz}), including also the dependence of the bias on the redshift in the latter case.  
The QML estimates are stable with respect to such different physical assumptions 
in the two fiducial power spectra. By considering the more physical scenario in which 
the bias evolves in redshift, the tension between 
the NVSS auto-spectrum and the theoretical predictions could be alleviated.

We note few important findings in our estimates of the angular power spectrum of the WMAP 7 year - NVSS cross-correlation. 
First, the estimate of the GG power spectrum in the NVSS map is {\em slightly} larger than our fiducial model. 
Our estimates for the NVSS auto-power spectrum agree very well with 
\cite{blakeferreiraborrill}, who used an optimal 
estimator similar to ours on a NVSS map of the same resolution of the one used here. 
The stability of the 
$C_\ell^{GG}$ estimates with respect to different flux threshold found in \cite{blakeferreiraborrill}
is also very similar to what we find.
 
As yet it is unclear whether this deviation could be caused by some systematics in the 
NVSS data or should be ascribed 
to a genuine physical effect, as an effective bias larger than $\sim 2$, 
which is usually assumed.
\cite{xia10a} estimated a larger discrepancy at lower multipoles and {explained} this effect as result of non-negligible primordial non-Gaussianity, caused by the large-scale scale-dependence of the non-Gaussian halo bias. 
However, the value inferred for the coupling non-Gaussian parameter $f_{NL}$ is much larger than the limits imposed by 
CMB analyses~\citep[e.g.,][]{komatsu11,curto}. The $f_{NL}$ constraints derived from the CMB-LSS 
cross-correlation~\citep{xia10b} provide lower values, in better agreement with the CMB tests. 
In addition, these authors also showed that when other LSS data sets are used~\citep[in particular, 
the QSOs sample of the SDSS][]{richards}, such non-Gaussian deviation is not found.


We also note an estimate lower than the fiducial model in the first bin of the $TG$ spectrum, 
less than {2$\sigma$} (as obtained by the Fisher matrix) lower than the fiducial model. This low value was not 
obtained in the previous investigation by \cite{Ho}.  



\begin{figure}
\includegraphics[width=8cm,angle=0]{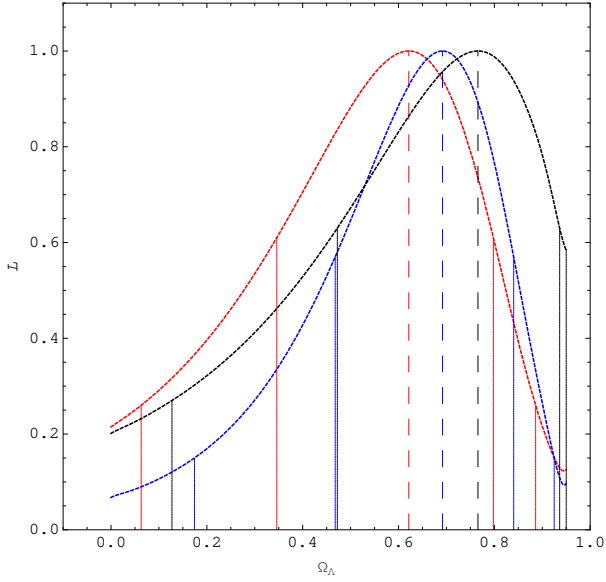}
\caption{The likelihood for ${\Omega}_{\Lambda}$ obtained by the Fisher
prescription in Eq. (\ref{fisher}), with the 95 \% and 68 \% C.L. for the threshold flux of 2.5 mJ (blue), 
5 mJ (red), 10 mJ (black) in NVSS, respectively.}
\label{likelihood_diffCuts_Noise}
\end{figure}

\begin{figure}
\includegraphics[width=8cm,angle=0]{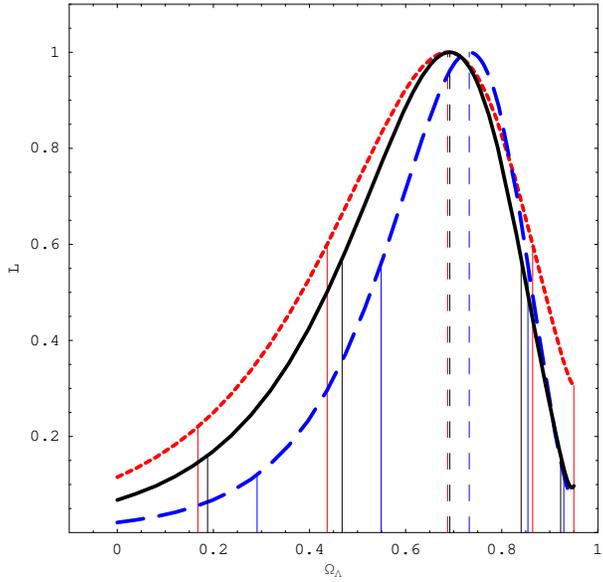}
\caption{Comparison of likelihood contours for ${\Omega}_{\Lambda}$ obtained by the Fisher prescription (black solid line),
the covariance computed by
the Monte Carlo for CMB only (red short dashed line) and for both CMB and LSS ({blue} dashed line).
The threshold flux in NVSS has been chosen to be 2.5 mJ.}
\label{likelihood_2p5mJ_MonteCarli}
\end{figure}

\begin{figure}
\includegraphics[width=8cm,angle=0]{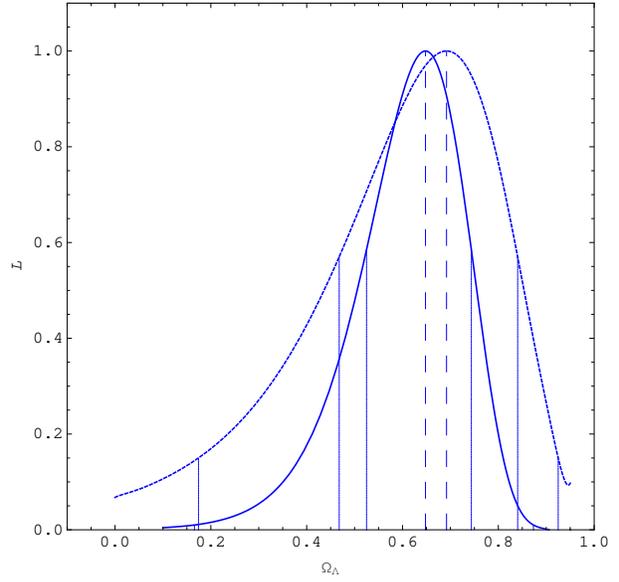}
\caption{Comparison of likelihood contours for ${\Omega}_{\Lambda}$ obtained by the Fisher
prescription in Eq. (\ref{fisher}) when accounting (dashed line) and when not accounting (solid line) for shot noise 
in NVSS data. 
The threshold flux in NVSS has been chosen as 2.5 mJ.}
\label{likelihood_2p5mJ_withANDwithoutNOISE}
\end{figure}

\begin{figure}
\includegraphics[width=8cm,angle=0]{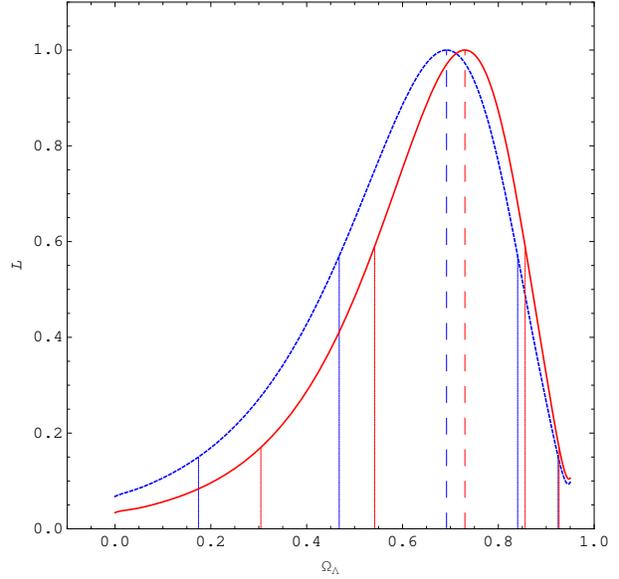}
\caption{Comparison of likelihood contours for ${\Omega}_{\Lambda}$ obtained by the Fisher
prescription in Eq. (\ref{fisher}) when considering the full covariance (red, solid line) and when using 
the approximation of a block diagonal signal covariance and Fisher matrix.
The threshold flux in NVSS has been chosen as 2.5 mJ.}
\label{likelihood_2p5mJ_div0}
\end{figure}

\section{Dark energy constraints}
\label{sect:constrain}

In this Section, we constrain the dark energy density $\Omega_{\Lambda}$ using the information 
contained in the ISW-LSS 
cross-correlation power spectrum, estimated through our QML.  
We assume the errors on the measured $C_{\ell}^{TG}$ are Gaussian, and calculate the relative likelihoods of $\Omega_\Lambda$ using  
\be
- 2 \ln[ {\cal{L}}(\Omega_\Lambda)] = \chi^2(\Omega_\Lambda)-\chi^2_{\rm min}.
\label{likelihood}
\ee
where 
\begin{eqnarray}
\label{chi2}
\chi^2(\Omega_\Lambda) =  \ \ \ \ \ \ \ \ \ \ \ \ \ \ \ \ \ \ \ \ \ \ \ \ \ \ \ \ \ \ \ \ \ \ \ \ \ \ \ \ \ \ \ \ \ \ \ \ \ \ \
 \ \ \ \ \ \ \ \  & \nonumber \\ 
\ \ \ \left[ C_{\ell}^{TG,{\rm obs}}-C_{\ell}^{TG}(\Omega_\Lambda) \right]
{\cal C}^{-1}_{\ell \ell'} \left[C_{\ell'}^{TG,{\rm obs}} -
C_{\ell'}^{TG}(\Omega_\Lambda) \right].&
\end{eqnarray}
Here $C_{\ell}^{TG, {\rm obs}}$ are the unbinned estimates of the cross-correlation power spectrum, 
{and} $C_{\ell}^{TG}(\Omega_\Lambda)$ are the theoretical predicted power spectrum.
The matrix ${\cal C}_{\ell \ell'}$ 
is the covariance matrix between different multipoles, 
which allows for correlations among non-diagonal terms which arise in the presence of masks.  
$\chi^2_{\rm min}$ is the minimum value of $\chi^2$ with respect to $\Omega_\Lambda$.

We compare the likelihoods obtained by different prescriptions for the covariance matrix. 
The first prescription is to use the {\em 
unbinned} QML estimates and the Fisher matrix as its covariance matrix:
\be
{\cal C}_{\ell \ell'}^F = (F^{-1})_{\ell \ell'}^{TG \, TG} \,.
\label{fisher}
\ee
An alternative prescription is to construct the covariance matrix ${\cal C}$ 
by {averaging over Monte Carlo realisations} of the maps.
For every model $\Omega_{\Lambda}$, we can define the covariance ${\cal C}$ with $N$ simulated CMB and LSS maps
\begin{eqnarray}
\label{cova}
{\cal C}_{\ell \ell'}(\Omega_{\Lambda}) =  \ \ \ \ \ \ \ \ \ \ \ \ \ \ \ \ \ \ \ \ \ \ \ \ \ \ \ \ \ \ \ \ \ \ \ \ \ \ \ \ \ \ \ \ \ \ \ \ \ \ \
 \ \ \ \ \ & \nonumber \\ 
\sum_{i=0}^{N} \frac{[ C_{\ell \ ,i}^{TG}(\Omega_{\Lambda})-\bar{C}_{\ell}^{TG}(\Omega_{\Lambda}) ] 
[ C_{\ell',i}^{TG}(\Omega_{\Lambda})-\bar{C}_{\ell'}^{TG}(\Omega_{\Lambda})]}{N},&
\end{eqnarray}
where the $C_{\ell \ ,i}^{TG}$ are the estimates for every single 
realization $i$ and the $\bar{C}_{\ell}^{TG}$ is their theoretical value.
However, we expect the covariance matrix not to depend strongly on the 
cosmological model, one can just consider the case with $\Omega_\Lambda=0$, 
and since $\bar{C}_{\ell}^{TG} (\Omega_{\Lambda}=0) =0$, the covariance becomes,
\be
{\cal C}_{\ell \ell'}^{MC} = \sum_{i=0}^{N} 
\frac{C_{\ell \ ,i}^{TG} C_{\ell',i}^{TG}}{N} .\,
\label{matr}
\ee

We can build ${\cal C}_{\ell \ell'}$ in Eq. (\ref{matr}) either by using random realisations of {only} the 
CMB maps {\sl and the single, true} NVSS map, or by creating a realisations of  
both CMB and LSS maps. In both cases, we generate our covariances on 1000 realisations, as done by 
\cite{vielva}. 
We also examine how the probability contours for $\Omega_\Lambda$ depend on the various assumptions such as the threshold flux cut used for the NVSS map or the sources redshift distribution.


We evaluate the likelihood with the various different prescriptions by sampling the $\chi^2$ on {values of 
$\Omega_{\Lambda}$, $ 0 < \Omega_{\Lambda} < 0.95 $.}
The other cosmological parameters are kept fixed to the values determined by WMAP~\citep{larson} for the standard $\Lambda$CDM model. 
As default NVSS description, the Eq. (\ref{honz}) model is
assumed, with a bias of 1.98.  

By adopting the Fisher matrix prescription in Eq. (\ref{fisher}), 
as tightest constraint  
we obtain $\Omega_\Lambda = 0.69^{+ 0.15 \, (0.23)}_{- 0.22 \, (0.50)}$ at $1 (2) \sigma$ 
confidence level (CL) for the lowest flux threshold of 2.5 mJ, see the {blue} dashed line in Fig. \ref{likelihood_diffCuts_Noise}. 
An Einstein-de Sitter Universe is disfavoured at more than 2 $\sigma$ CL for the lowest flux threshold in NVSS, consistent with earlier measurements. 
Note that the conditional probabilities for $\Omega_\Lambda$ agree for the different flux thresholds considered.

By building the covariance through realizations of the CMB maps while keeping the NVSS map fixed, we obtain the probability distribution given by the {red} dashed line of Fig. \ref{likelihood_2p5mJ_MonteCarli}.   
We find $\Omega_\Lambda = 0.69^{+0.18 \, (0.26)}_{-0.25 \, (0.52)}$ at $1 (2) \sigma$ CL.
Using instead the covariance derived from realizations of both CMB and LSS {maps}, the probability distribution given by the blue dashed line
of Fig. \ref{likelihood_2p5mJ_MonteCarli} we find $\Omega_\Lambda = 0.73^{+0.12 \, (0.18)}_{-0.20 \, (0.44)}$ at $1 (2) \sigma$ CL.
Note that the constraint based on the Fisher covariance is tighter than the one based on a Montecarlo covariance keeping
fixed the NVSS map, but looser than the Montecarlo covariance obtained with CMB and LSS uncorrelated maps. 
Overall, the three likelihood prescriptions are consistent, 
although 
some of the differences might be ascribed to the (unexplained) discrepancy between the 
$C_{\ell}^{GG}$ estimates and the theoretical predictions, on which the Montecarlos are based.

Given the agreement among the three different likelihood prescriptions, we can use the Fisher prescription for the covariance to  
test other dependences of the analysis. In Fig. (\ref{likelihood_2p5mJ_withANDwithoutNOISE}) we verify the importance 
of taking into account the shot noise in the NVSS map: by not taking into account the shot noise the probability contours for 
$\Omega_\Lambda$ would be much tighter, even for the maps with the most sources.  
We then study the impact of approximating the signal covariance matrix (and consequently the 
Fisher matrix) as block diagonal, i.e. considering 
$C_{\ell}^{TG}=0$ for the fiducial underlying model. This approximation is not essential for 
our approach, whereas it is necessary for 
\cite{padmanabhan} in which only $C_{\ell}^{TG}$ is estimated.
As mentioned in Sect. 2.3, the difference in the power spectrum 
estimates is barely visible and we have verified by MonteCarlo that 
this approximation does not alter the optimality of the method. On the real data considered here, Fig. (\ref{likelihood_2p5mJ_div0}) 
shows how the constraints with the full covariance and Fisher are a bit tighter than those with the block diagonal assumption. 
As already noticed in \cite{gruppusowmap5}, conditional probability slices are much more sensitive 
to small changes than QML estimates.

In Fig. (\ref{likelihood_2p5mJ_deZottiVSHo}) we compare the redshift distribution estimated 
with CENSORS data by \cite{dezottireview} in Eq. (\ref{dezottinz}), including a redshift dependent bias as from Eq. (),  
with the one adopted by \cite{Ho}, with an effective bias of $1.98$.

\section{Discussions and Conclusions}
\label{sect:concl}

We have developed an optimal estimator for the angular power spectrum of the cross-correlation 
between CMB and maps of large scale structure, which in parallel estimates their auto-spectra. 
This has been tested using an ensemble of randomly generated maps, 
and we have demonstrated the robustness of the QML estimates 
for the TT, TG and GG power spectra. Our QML implementation extend similar optimal estimators 
limited only to the galaxy auto power spectrum \citep{blakeferreiraborrill} 
or only to the cross-correlation power spectrum \citep{padmanabhan}.

We have applied our method to WMAP 7 year and NVSS data, the best public data sets at present for studying the ISW 
cross-correlations.  Our method makes no assumptions, and allows to measure the cross-correlation with optimal errors 
and to exploit the full cosmological information contained in the maps, though our analysis is limited to a pixel resolution of$1.8^\circ$. 
While the NVSS map contains known declination systematics, we correct for these and find, as has earlier work, that they appear to have little effect on the measured cross-correlations.  In agreement with previous studies, we detect a non-zero cross-correlation, and have also seen a slight excess in the NVSS auto-angular power spectrum compared to that expected theoretically.   

We have translated these measurements into the quantitative constraints on the fraction of dark energy in a $\Lambda$CDM model which can be obtained only by the cross-correlation of WMAP and NVSS, estimating  $\Omega_\Lambda$ while keeping fixed all the other cosmological parameters to the WMAP 7 yr best-fit values \citep{larson}.
We have compared three different prescriptions for estimating the covariances: using the Fisher matrix computed by our QML,
on {Monte Carlo} relisations of the CMB maps only and creating {Monte Carlo} realisations of both CMB and LSS maps.
We have found a good agreement among the $\Omega_\Lambda$ probability contours obtained from these three different likelihood prescriptions. The width of this probability contour depends mainly on the flux threshold and associated level of Poisson noise in the NVSS map, but the signal amplitude seems largely independent of the flux. 
The constraint from the likelihood prescription based on the Fisher matrix we derive from the cross-correlation between WMAP 7 yr and NVSS 
data is 
$\Omega_\Lambda = 0.69^{+ 0.15 \, (0.23)}_{- 0.22 \, (0.52)}$ at $1 (2) \sigma$
confidence level (CL) for the lowest flux threshold of $2.5$ mJ. Such value is quite consistent with the concordance cosmology
This result agrees with that expected from a typical survey with sky fraction and noise property as the 
NVSS, and agrees with \cite{vielva}, but  
is somewhat weaker than the one obtained by the non-optimal analysis by \cite{PBM} 
based on needlets. 
It is not clear if this 
discrepancy is due to the 
lower resolution considered here or the neglection of shot 
noise in the NVSS map in the analysis by \cite{PBM}.  

\section*{Acknowledgements}

We wish to thank G. de Zotti for useful discussions on NVSS and S. Matarrese for 
stimulating discussions on the bias.
FS, FF and AG wish to thank Adriano De Rosa for useful suggestions on the QML implementation.
We acknowledge the use of the SP6 supercomputer at CINECA under the 
agreement INAF/CINECA and LFI/CINECA, and the use of the HEALPix 
software and analysis package (\cite{gorski}). 
We also acknowledge the use of the Legacy Archive for Microwave 
Background Data Analysis (LAMBDA); support for LAMBDA is provided by the NASA Office of Space Science. 
This work is supported by ASI through ASI/INAF Agreement I/072/09/0 for the Planck LFI Activity of Phase E2. 
PV, RBB, AMC and EMG acknowledge partial financial support from 
the Spanish Ministerio de Ciencia e Innovaci\'on project AYA2010-21766-C03-01 
and the Consolider Ingenio-2010 Programme project CDS2010-00064, and PV also 
acknowledges financial support from the Ram\'on y Cajal programme. RC is supported by STFC grant ST/H002774/1.

\end{document}